\documentclass[fleqn,usenatbib]{mnras}

\usepackage{newtxtext,newtxmath}

\usepackage[T1]{fontenc}
\usepackage{longtable}
\DeclareRobustCommand{\VAN}[3]{#2}
\let\VANthebibliography\thebibliography
\def\thebibliography{\DeclareRobustCommand{\VAN}[3]{##3}\VANthebibliography}

\usepackage{graphicx}	
\usepackage{amsmath}	

\title[Clustering of stars around IL Cep]{Clustering of low mass stars around Herbig Be star IL Cep - Evidence of "Rocket Effect" using {\textit Gaia} EDR3 ?}

\author[R. Arun et al.]{
R. Arun,$^{1}$\thanks{E-mail: arun.roy@res.christuniversity.in}
Blesson Mathew,$^{1}$
G. Maheswar,$^{2}$
Tapas Baug,$^{3}$
Sreeja S. Kartha,$^{1}$
G. Selvakumar,$^{2}$
P. Manoj,$^{4}$
\newauthor
B. Shridharan,$^{1}$
R. Anusha,$^{1}$
Mayank Narang,$^{4}$ 
\\
$^{1}$Department of Physics and Electronics, CHRIST (Deemed to be University), Bangalore 560029, India\\
$^{2}$Indian Institute of Astrophysics, Sarjapur Road, Koramangala, Bangalore 560034, India\\
$^{3}$S.N. Bose National Centre for Basic Sciences,Block-JD, Sector-III, Salt Lake, Kolkata-700106, India\\
$^{4}$Tata Institute of Fundamental Research, Homi Bhabha Road, Mumbai 400005, India
}

\date{Accepted XXX. Received YYY; in original form ZZZ}

\pubyear{2015}

\begin{document}
\label{firstpage}
\pagerange{\pageref{firstpage}--\pageref{lastpage}}
\maketitle

\begin{abstract}
We study the formation and the kinematic evolution of the early type Herbig Be star IL Cep and its environment. The young star is a member of the Cep OB3 association, at a distance of 798$\pm$9 pc, and has a "cavity" associated with it. We found that the B0V star HD 216658, which is astrometrically associated with IL Cep, is at the center of the cavity. From the evaluation of various pressure components created by HD 216658, it is established that the star is capable of creating the cavity. We identified 79 co-moving stars of IL Cep at 2 pc radius from the analysis of {\textit Gaia} EDR3 astrometry. The transverse velocity analysis of the co-moving stars shows that they belong to two different populations associated with IL Cep and HD 216658, respectively. Further analysis confirms that all the stars in the IL Cep population are mostly coeval ($\sim$ 0.1 Myr). Infrared photometry revealed that there are 26 Class II objects among the co-moving stars. The stars without circumstellar disk (Class III) are 65\% of all the co-moving stars. There are 9 intense H$\alpha$ emission candidates identified among the co-moving stars using IPHAS H$\alpha$ narrow-band photometry. The dendrogram analysis on the Hydrogen column density map identified 11 molecular clump structures on the expanding cavity around IL Cep, making it an active star-forming region. The formation of the IL Cep stellar group due to the "rocket effect" by HD 216658 is discussed.      
\end{abstract}




\section{Introduction}
\label{sect:into}

 
 Infrared bubbles or cavity-like structures are ubiquitous in the Milky way \citep{Churchwell2006ApJ...649..759C, Churchwell2007ApJ...670..428C}. The formation of the structures is due to the combined effects of radiation \citep{Bisbas2011IAUS..270..263B}, ionization \citep{Sternberg2003ApJ...599.1333S} and stellar winds \citep{Dale2015MNRAS.450.1199D} from one or more massive central stars ($\geq$ 8 M\textsubscript{\(\odot\)}; \citealp{Zinnecker2007ARA&A..45..481Z}). The cavities are generally associated with infrared bright rims with high star-forming activity \citep{Morgan2004A&A...426..535M}. The triggered star formation in the region can create another generation of massive young stars. \cite{Fuente2002A&A...387..977F} identified cavity structures associated with a sample of Herbig Ae/Be (HAeBe) stars. Thus HAeBe stars and their associated regions are ideal for studying various aspects of triggered star formation.

HAeBe stars are intermediate-mass (2--10 M\textsubscript{\(\odot\)}) pre-main sequence (PMS) stars with a circumstellar accretion disk \citep{Herbig1960,WATERS1998}. The dust and gas content in the circumstellar disk/medium produce infrared excess in the spectral energy distribution of HAeBe stars. The recombination emission lines of Hydrogen, Calcium, Iron, etc., are also formed in the disk \citep{Hamann1992}. From the near-infrared (NIR) imaging studies, \cite{Testi1997, Test1998} identified clustering of low mass young stellar objects (YSO) with HAeBe stars. Also, the spatial density concentration of the low mass YSOs is correlated with the spectral type of the HAeBe star in the center of the distribution \citep{Test1999}. The YSOs around HAeBe stars are astrometrically associated with the central HAeBe stars \citep{Saha2020}.

IL Cep (also called HD 216629), a B3 spectral type \citep{Merrill1949ApJ...110..387M} and a confirmed PMS star \citep{Assousa1977ApJ...218L..13A, The1994A&AS..104..315T} belongs to Cep OB3 association \citep{Blaauw1959ApJ...130...69B, Garmany1973AJ.....78..185G}. The star is reported to be associated with the reflection nebula GN 22.51.3 \citep{Magakian2003A&A...399..141M}. IL Cep belongs to a visual binary system with a companion star HD 216629B at a separation of 7" \citep{MelNikov1996ARep...40..350M}. Several studies indicate that the star IL Cep itself is an unresolved binary star system consisting of stars having spectral types B3(±2) and B4(±2) \citep{Wheel2010, Ismailov2016A}. From NIR photometric analysis \cite{Testi1997} found clustering of 24 low mass stars with IL Cep. Later, \cite{Fuente2002A&A...387..977F} reported cavity formation around IL Cep from the analysis of \textsuperscript{13}CO and C\textsuperscript{18}O data. The presence of the cavity is reconfirmed by \cite{Zhang2016MNRAS.458.4222Z}. They used \textsuperscript{13}CO $(J = 1-0)$ continuum data and Wide-field Infrared Survey Explorer (WISE) \citep{Cutri2013} images to define the bounds of the cavity around the IL Cep region. They argue that the violent past of IL Cep created the cavity.  \cite{Zhang2016MNRAS.458.4222Z} proposed that the center of the cavity is occupied not by IL Cep but by a slightly more massive star HD 216658. Interestingly, \cite{Zhang2016MNRAS.458.4222Z} found that the cavity is not formed due to HD216658. The cavity formation is due to the mass dispersal process triggered by various pressure components of the massive central star in the region. The cavity formation considerably reduces the line of sight extinction towards the region, which gives a unique opportunity to study the clustering using optical {\textit Gaia} EDR3 data.

The identification of clustering around HAeBe star is not investigated from an astrometric perspective thoroughly. The study carried out by \cite{Saha2020} in the case of the star HD 200775 has shown that 80\% of the astrometrically identified co-moving stars are diskless stars (Class III), and spectroscopy of four of them found to show weak H$\alpha$ emission. As a result, these stars will be missed in NIR and H$\alpha$ surveys which are classically used to identify the PMS populations. We are using the {\textit Gaia} Early Data Release 3 ({\textit Gaia} EDR3) to find the young co-moving stars around IL Cep, thereby identifying clustering of low mass YSOs around Herbig Be stars. Also, we combined infrared photometry and spectroscopic analysis to identify possible disk-bearing stars among the co-moving stars.

\begin{figure*}
   \centering
   \includegraphics[width=1\columnwidth]{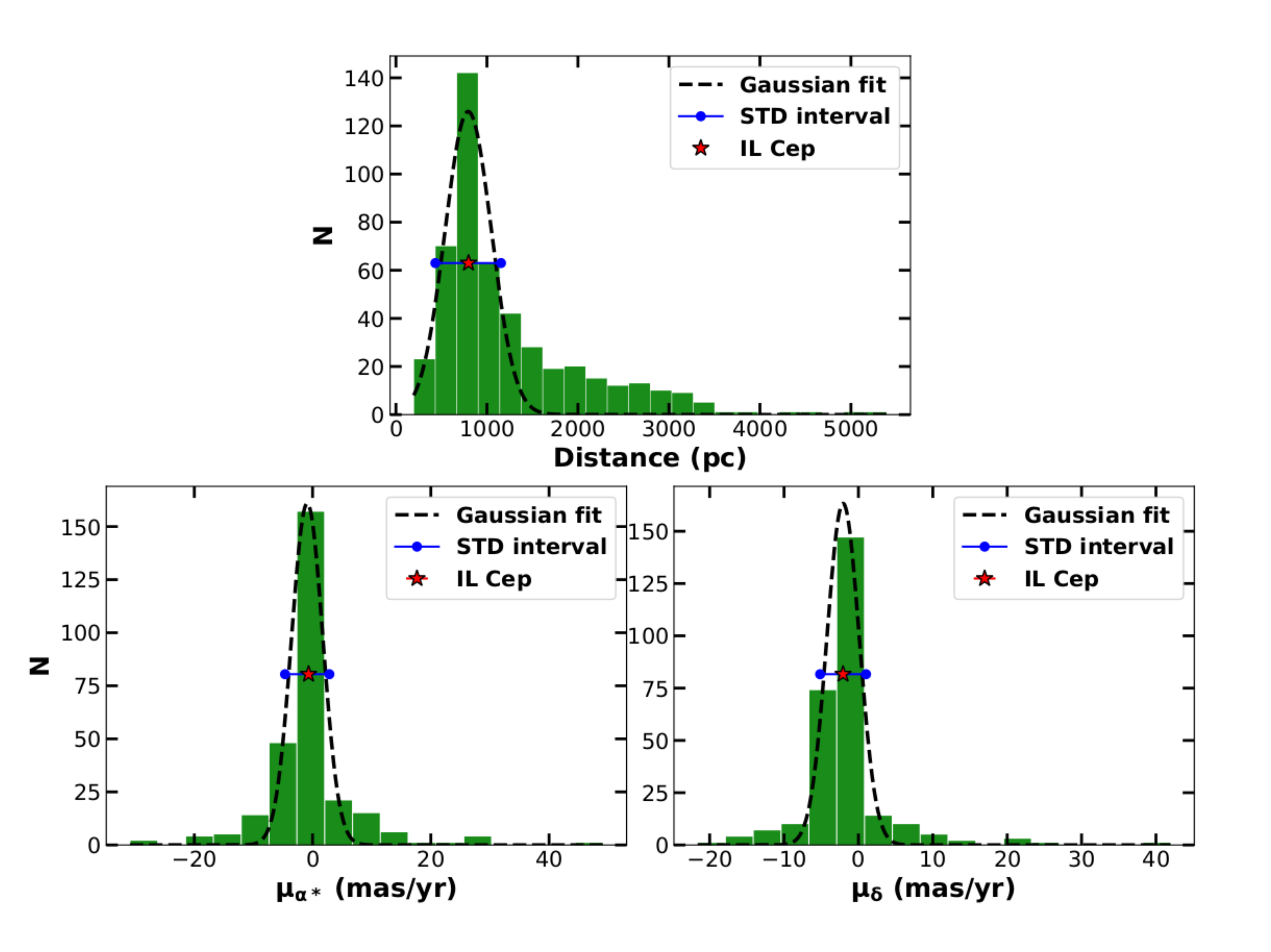}
   \includegraphics[width=1\columnwidth]{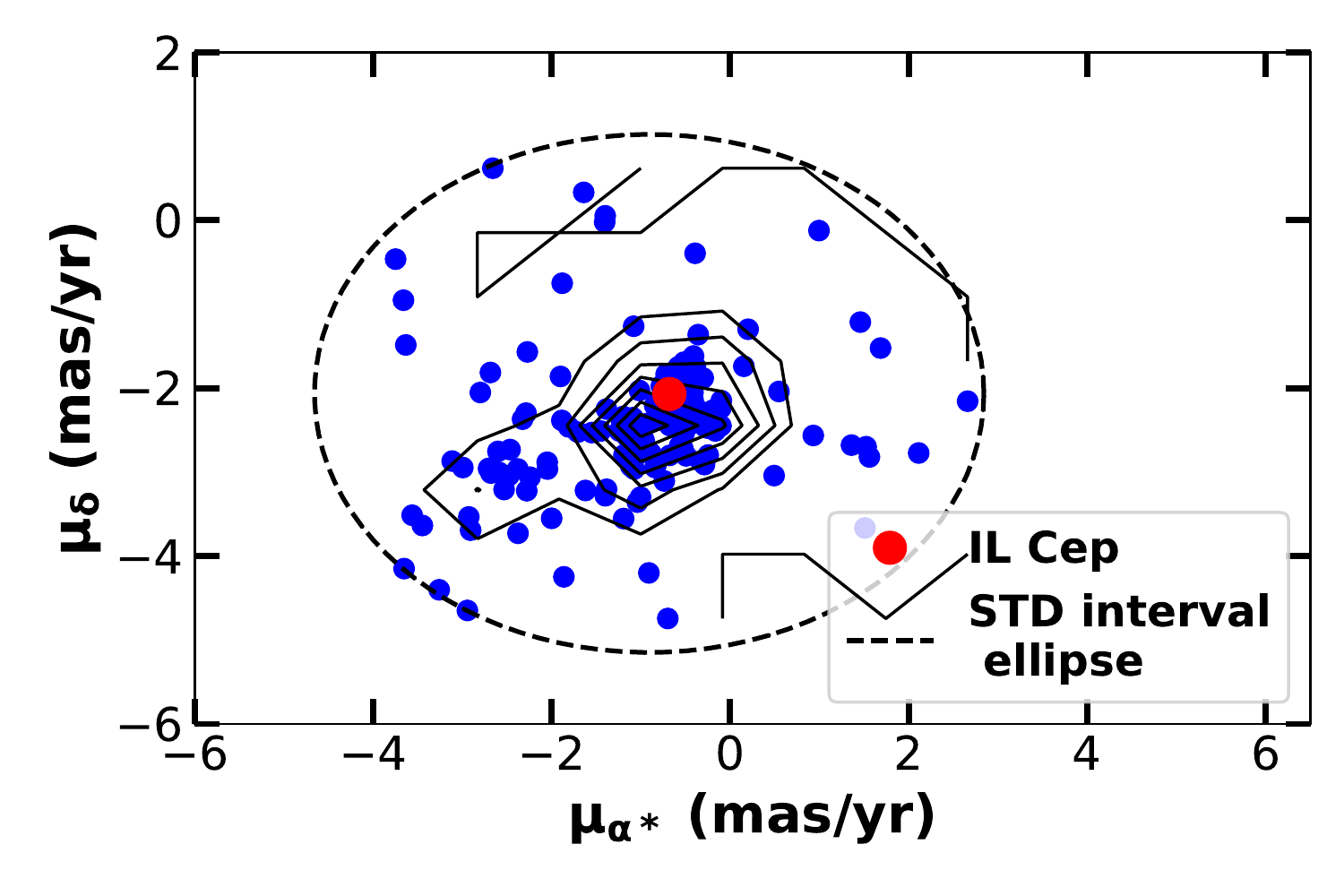}
   \caption{The figures show the Gaussian fits on the histograms of distance and proper motion value of stars inside a 2 pc radius around IL Cep and the vector point diagram of "level one" sources. Using the Gaussian fits, we calculated standard deviation intervals for each parameter. IL Cep appears to be inside all three standard deviation intervals, indicating an overdensity of stars around IL Cep. The stars that satisfy the standard deviation intervals are taken as "level one" samples for further analysis. In the vector point diagram, the dashed ellipse is created using the standard deviation intervals of proper motion values. The black contours are the number density contours that indicate an overdensity of stars near IL Cep.}

   \label{Fig1}
   \end{figure*}

The paper is organized as follows. The data used in this study are described in Sect. \ref{sect:data}. Sect. \ref{sect:results} explains various analyses carried out for identifying the clustering around IL Cep, identification of disk-bearing stars, and detection of molecular clumps in the region. Also, the possibility of the rocket effect being the initial trigger for the formation of IL Cep and the co-moving stars is given in Sect. \ref{sect:results}, and the significant results are summarized in Sect. \ref{sect:conclusion}.

\section{Data Inventory}
\label{sect:data}

The {\textit Gaia} EDR3, made available on 3\textsuperscript{rd} December 2020, contains astrometric and photometric data of 1.8 billion sources \citep{EDR3_I,Lindegren2020arXiv201203380L}. The astrometric data and three photometric band magnitudes of the stars in the region around IL Cep are taken from {\textit Gaia} EDR3. The direct conversion of {\textit Gaia} parallax values has its inherent problems (see \citealp{bailerjohns2018} for a detailed discussion). \cite{Bailer2020} estimated the distances of 1.47 billion {\textit Gaia} EDR3 sources using two methods. The first one is the geometric distance which uses distance likelihood (from {\textit Gaia} parallax) and a distance prior (an exponentially decreasing space density prior that is based on a Galaxy model) approach, which was also applied in the {\textit Gaia} DR2 \citep{bailerjohns2018}. The second approach is photogeometric, which uses photometric color and apparent magnitude along with the {\textit Gaia} parallax. Both the distances show no considerable differences in our sample. Thus we used the geometric distance from \cite{Bailer2020} for all the stars studied in this work.

The NIR magnitudes used in the analysis are 2MASS J, H, and K\textsubscript{S}  \citep{Skrutskie2006}. Also, the mid-infrared (MIR) magnitudes, IRAC [3.6] and IRAC [4.5], are taken from \textit{Spitzer} Glimpse 360 \citep{Whitney2011AAS...21724116W}. We also used the Isaac Newton Telescope (INT) Photometric H-Alpha Survey (IPHAS) \citep{Drew2005MNRAS.362..753D, Barentsen2011MNRAS.415..103B} magnitudes to identify H$\alpha$ sources. 

The low-resolution optical spectrum of HD 216658, the massive star in the region, and HD 216629B, the visual binary companion of IL Cep, were obtained using the Optomechanics Research (OMR) spectrograph \citep{Prabhu1998BASI...26..383P} mounted on the 2.34 m Vainu Bappu Telescope (VBT), Vainu Bappu Observatory (VBO), Kavalur. The spectra are obtained using the grating centered at the H$\alpha$ line at 6563 \AA. The resolution of the observed spectrum is about 8 \AA. The dome flats are obtained for flat fielding the images. The bias subtraction, flat field correction, and spectral extraction are performed with standard IRAF tasks. The wavelength calibration of the spectra was carried out using spectra of FeNe and FeAr calibration lamps. All the extracted raw spectra were wavelength calibrated and continuum normalized using IRAF tasks. For all the stars the average signal-to-noise ratio (SNR) near H$\alpha$ is above 100. The log of observations of the stars is shown in the \autoref{tab:log}. 

The \textsuperscript{12}CO (\textit{J = 1$-$0)} observations of the IL Cep region were retrieved from the Canadian Astronomy Data Centre (CADC). The data is a part of the Canadian Galactic Plane Survey (CGPS) and observed in 1995 using the Five College Radio Astronomy Observatory (FCRAO) 14 m telescope. The data was initially part of the FCRAO Outer Galaxy Survey (OGS; \citealp{Heyer1998ApJS..115..241H}). The data has a spatial resolution of 45" and a velocity resolution of 0.15 km~s\textsuperscript{-1}. The noise level per channel is 0.75 K (T\textsuperscript{*}\textsubscript{R}).

\begin{table}
\centering
\caption{The table provides log of observation of HD 216658 and IL Cep B}
\label{tab:log}
\begin{tabular}{ccccc}
\hline
Star & \begin{tabular}[c]{@{}c@{}}V \\ (mag)\end{tabular} & \begin{tabular}[c]{@{}c@{}}Spectral \\ type\end{tabular} & \begin{tabular}[c]{@{}c@{}}Date of\\ observation\end{tabular} & \begin{tabular}[c]{@{}c@{}}Exposure time\\ (s)\end{tabular} \\ \hline
HD 216658 & 8.9 & B0.5V & 12-12-2020 & 1200 \\
IL Cep B & 13.8 & A6V & 26-12-2020 & 2400 \\ \hline
\end{tabular}
\end{table}

\section{Results}
\label{sect:results}

 \begin{figure}
   \centering
   \includegraphics[width=\columnwidth]{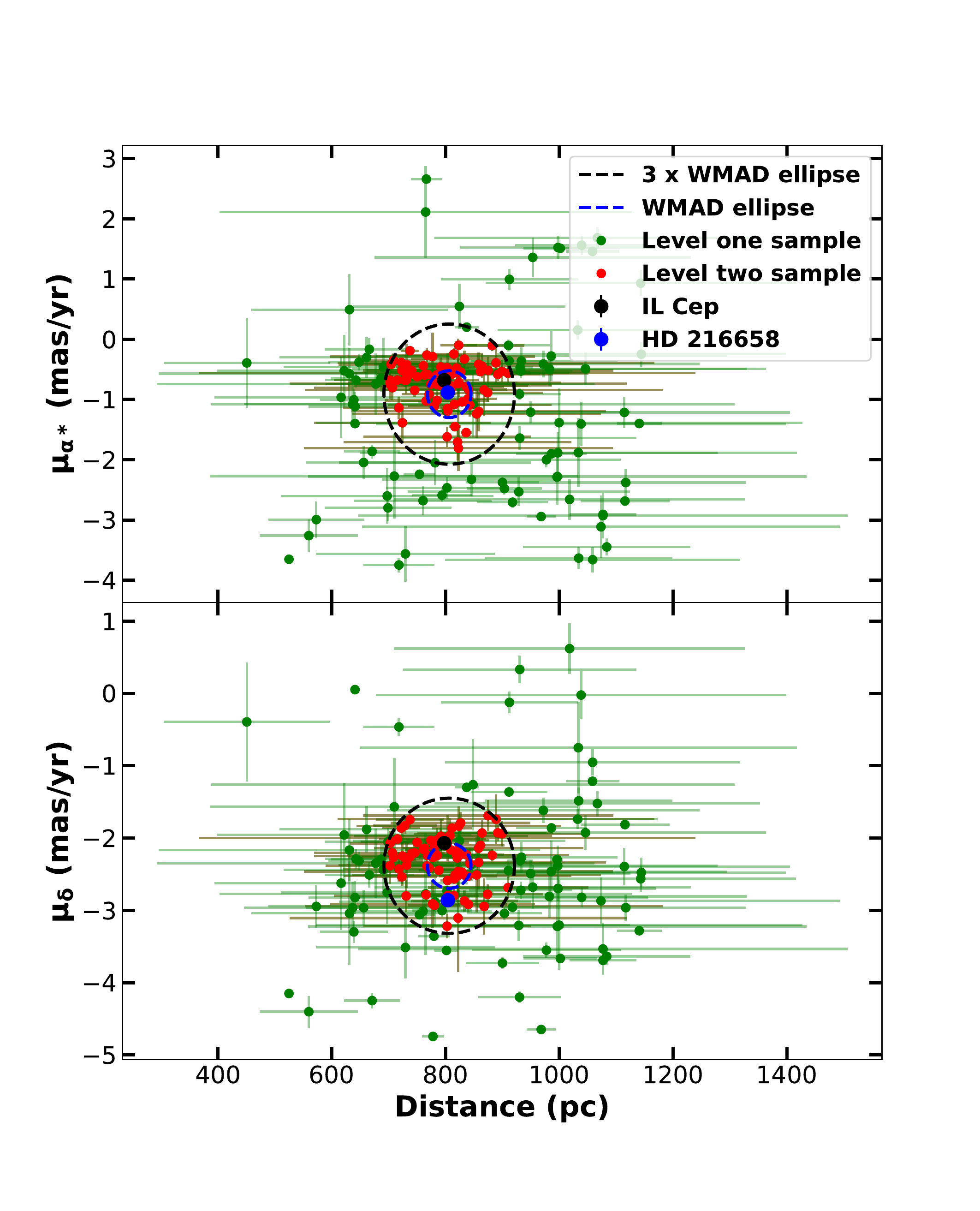}
   \caption{The figures show the distribution of selected co-moving stars of IL Cep in d vs $\mu_{\alpha *}$ and d vs $\mu_{\delta}$ diagrams. The selected stars are shown using red color star symbols and rejected stars are shown in green. The WM and 3 $\times$ WMAD ellipse are shown in blue and black dashes curves, respectively. IL Cep and HD 216658 are also shown in the figure. }
   \label{Fig2}
   \end{figure}

\subsection{{\textit Gaia} EDR3 astrometric analysis}\label{sec:astrometry}
 In this study, we use the {\textit Gaia} EDR3 for our astrometric analysis \citep{EDR3_I}. The star IL Cep is at a distance of 798$^{+8}_{-10}$ pc \citep{Bailer2020} and is part of the Cep OB3 association \citep{Blaauw1959ApJ...130...69B, Garmany1973AJ.....78..185G}. Clustering of low mass YSOs around the Herbig Be star IL Cep was studied by \cite{Test1998} using infrared K band observations. 
 
 We selected all {\textit Gaia} EDR3 detections around IL Cep within a radius of 8.5' (\textasciitilde 2 pc) and found 1565 sources. We chose 2 pc as our search radius because the core radius of an open cluster is 1-2 pc \citep{Moraux2016EAS....80...73M}. This search radius is sufficient to identify clustering, if any, around IL Cep. The detections within our target field are cross-matched with the distance catalog of \cite{Bailer2020} using EDR3 source ID. We found 1366 matches with distances from  \cite{Bailer2020}. Further, the stars with good quality astrometry are selected using the re-normalized unit weight error (RUWE) parameter and an uncertainty cut using the parallax values \citep{Saha2020}. The stars with RUWE > 1.4 are avoided due to the {\textit Gaia} EDR3 astrometry quality recommendations \citep{Fabricius2020arXiv201206242F}. And finally, the stars with parallax more than 3-sigma confidence (i.e. parallax/error in parallax >
3) are selected for the analysis. We compiled 477 stars including IL Cep in the 8.5' radius that satisfies the defined astrometric quality criterion as our initial sample for further analysis. 
 
 The stars clustered together have been identified by overdensity in the distribution of astrometric parameters \citep{Castro2020A&A...635A..45C}. We used gaussian fitting method on proper motion in right ascension ($\mu_{\alpha *}$ = $\mu_{\alpha}cos\delta$), proper motion in declination ($\mu_{\delta}$) and distance (d) histograms to check for overdensity of stars around IL Cep. Each of the three histograms showed a single Gaussian distribution. We fitted these histograms with a Gaussian function. The histograms and the fit are shown in \autoref{Fig1} (left). The three Gaussian fits provided standard deviation intervals for the three parameters. The standard deviation intervals are shown in the \autoref{Fig1} (left). The Gaussian fitting statistics are given in \autoref{tab:gauss}.
 
\begin{table}
\centering
\caption{The table gives the Gaussian fitting statistics of histograms shown in \autoref{Fig1} (left).}
\label{tab:gauss}
\begin{tabular}{cccc}
\hline
\begin{tabular}[c]{@{}c@{}}Astrometric\\ parameter\end{tabular} & Mean & Amplitude & \begin{tabular}[c]{@{}c@{}}Standard\\ Deviation\end{tabular} \\ \hline
Distance (pc) & 794.0 & 126.0 & 360.5 \\
$\mu_{\alpha *}$ (mas yr\textsuperscript{-1}) & -0.91 & 161.0 & 3.75 \\
$\mu_{\delta}$ (mas yr\textsuperscript{-1}) & -2.06 & 163.3 & 3.08 \\ \hline
\end{tabular}
\end{table}

 IL Cep appears to have an overdensity of stars around it because it is inside the standard deviation intervals of all three parameters. We selected the 165 stars that are inside the standard deviation intervals of all three astrometric parameters. This sample of stars is called the "level one" sample from now. \autoref{Fig1} (right) shows the vector point diagram (VPD) of the "level one" sample along with IL Cep. The ellipse shown in the VPD is made using the standard deviation intervals. The number density contours illustrate the overdensity of stars around IL Cep. From the figure, it is clear that there are stars in the "level one" sample that may not be associated with IL Cep. Thus we used the "level one" sample to constrain the astrometric parameters further using a median and median absolute deviation method adopted by \cite{Saha2020}. The Gaussian fitting analysis helped us in removing outliers and provided a sample of 165 stars for further analysis. We updated both the parameters to the weighted median (WM) and the weighted median absolute deviation (WMAD). The weights are decided using {\textit Gaia} G magnitude brightness of the stars. The "weighted" median criterion is adopted due to the increase in the uncertainty of {\textit Gaia} astrometry with the decrease in the brightness of stars ({\textit Gaia} Data Release 2 - Documentation release 1.2)\footnote{https://gea.esac.esa.int/archive/documentation/GDR2/}.
 
 \begin{figure*}
   \centering
   \includegraphics[width=1.3\columnwidth]{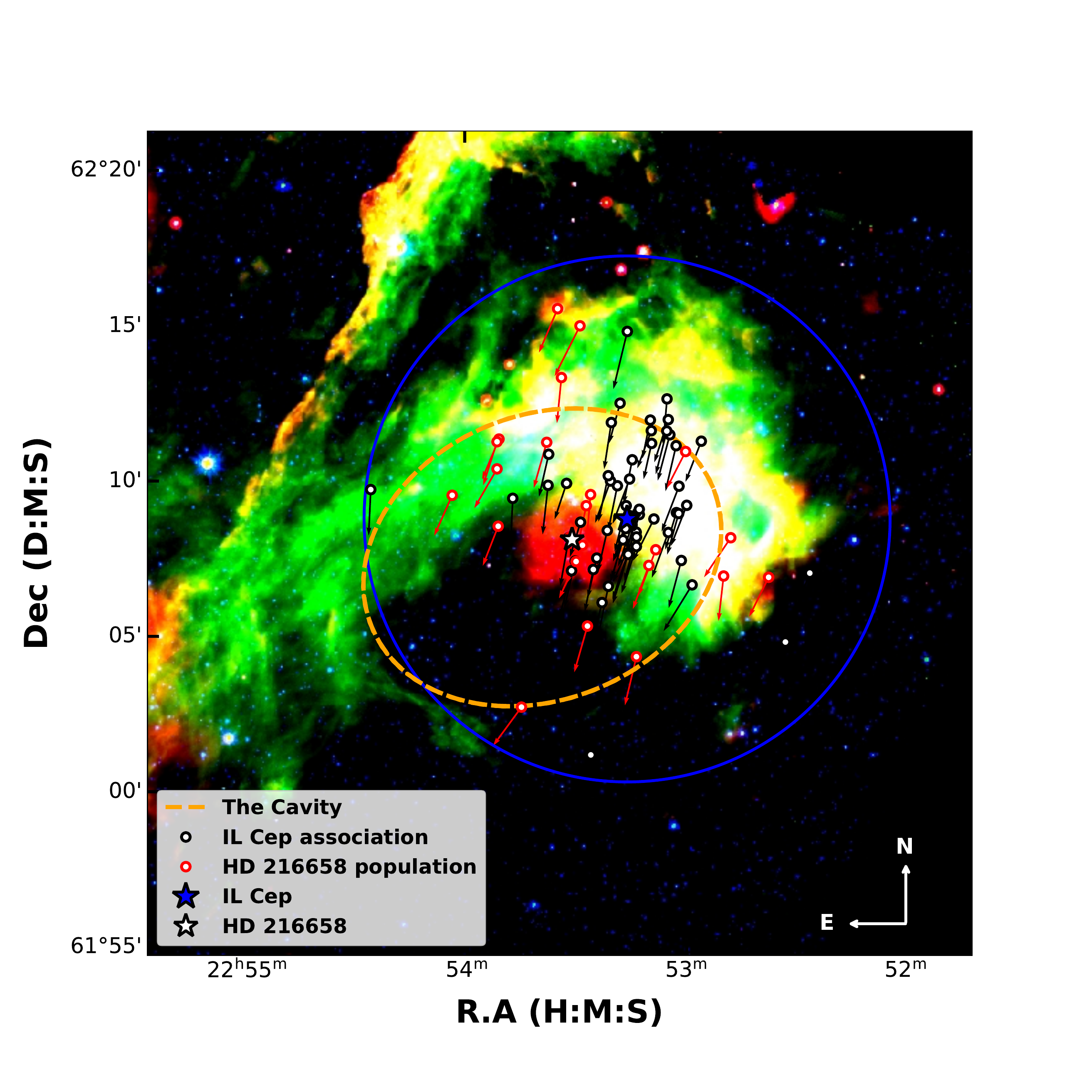}
\caption{The RGB images of the 25$'$ $\times$ 25$'$ region around IL Cep using \textit{Spitzer} (R = MIPS 24 $\mu m$, G = IRAC 8 $\mu m$, and B = IRAC [3.6] $\mu m$ ) is shown. The brightest star in the region, HD 216658 and IL Cep are marked in the figure. The co-moving stars associated with IL Cep are shown in black encircled symbols and the population associated with HD 216658 is shown in red encircled symbols. The "cavity" defined by \protect\cite{Zhang2016MNRAS.458.4222Z} is shown with an dashed orange ellipse and the 2 pc search radius is shown as a blue circle.}
   \label{Fig3}
\end{figure*} 

 The WM and WMAD of d are estimated as 806 pc and 38 pc respectively. The WM and WMAD of $\mu_{\alpha *}$ is estimated to be -0.91 mas yr\textsuperscript{-1} and 0.39 mas yr\textsuperscript{-1} respectively. The same for $\mu_{\delta}$ is given as -2.38 mas yr\textsuperscript{-1} and 0.31 mas yr\textsuperscript{-1} respectively. Two ellipses are defined using WM and 3 $\times$ WMAD values of d vs $\mu_{\alpha *}$ and d vs $\mu_{\delta}$. The stars that are inside the ellipses are selected as the associated stars of IL Cep (here onwards "level two" sample). \autoref{Fig2} shows the d vs $\mu_{\alpha *}$ and d vs $\mu_{\delta}$ plot with required ellipses. The level one and level two samples are illustrated in the figure.  We identified 78 stars from the astrometric analysis, which are clustered around IL Cep inside a 2 pc radius in the sky plane. \autoref{Fig3} shows the RGB image of the region made using MIPS 24 $\mu m$, IRAC 8 $\mu m$, and IRAC 3.6 $\mu m$ images. The newly identified co-moving stars, IL Cep, and the brightest star in the region, HD 216658 are marked in the \textit{Spitzer} color composite image with the proper motion vectors embedded on them. The proper motion vectors suggest that the selected stars are having similar motion through the sky. The astrometric and photometric data of all 78 co-moving stars are listed in the \autoref{tab:appendix_1} 
 
  The radius of the region we adopted is 8.5'. To check the possibility of having more co-moving stars in the extended region which is part of the larger star-forming cloud Cep OB3, we checked for stars with similar astrometric constraints in another region near IL Cep. To verify this, we took 8.5' regions 1\textdegree~ south of IL Cep. The number of stars that satisfy the WM and 3 $\times$ WMAD ellipse constraints in the southern region is 5. This shows that even though there may be similar moving stars in the extended regions, there is a clear overdensity of stars that satisfy the astrometric constraints using WM and 3 $\times$ WMAD around IL Cep. The "cavity" defined by \cite{Zhang2016MNRAS.458.4222Z} has a sky area of 0.05 square degrees and the search radius we used has an area of 0.063 square degrees. Both cavity and search radius are shown in \autoref{Fig3}. The cavity contains 77\% of all the co-moving stars, .i.e. 60 out of 78 stars are inside the "cavity". This directly implies that there are co-moving stars clustered around IL Cep and the over-density is associated with the cavity. 
  
  To summarize, we compiled the stars with high quality astrometric data around IL Cep. Using Gaussian analysis, we filtered out layers and retained 165 stars as level one sample. Finally, using WM and 3 $\times$ WMAD values of the sample we identified a total of 78 co-moving stars of IL Cep.

\subsection{HD 216658 - a massive star astrometrically associated with IL Cep and its co-moving stars}\label{sec:HD 216658_I}

The brightest star in the region of our study is HD 216658, with a V magnitude of 8.9 mag \citep{Brodskaya1953IzKry..10..104B}. IL Cep is slightly fainter with a V magnitude of 9.36 mag \citep{Hog2000A&A...355L..27H}. From the literature, it is identified that HD 216658 is of B0-0.5V spectral type \citep{Morgan1953ApJ...118..318M, Garrison1970AJ.....75.1001G}, and IL Cep is of B3 spectral type \citep{Merrill1949ApJ...110..387M}. \cite{Zhang2016MNRAS.458.4222Z} suggested that HD 216658 is geometrically positioned at the center of the cavity, but IL Cep is the predominant exciting star in the region. This is quite surprising since HD 216658 is more massive than IL Cep, and hence, should emit more UV flux that can trigger more mass dispersal and create a cavity. We are investigating the possibility of HD 216658 being the cause of cavity formation in the region. From {\textit Gaia} DR2 estimates HD 216658 is a foreground star at a distance of 668$^{+43}_{-38}$ pc. This is at a distance of 130 pc from IL Cep, which was reported at 798$^{+18}_{-17}$ pc \citep{bailerjohns2018}. The RUWE parameter of HD 216658 according to {\textit Gaia} DR2 data is 2.4, which is significantly higher than the quality cut we gave for the astrometric analysis in Sect. \ref{sec:astrometry}. \autoref{tab:stellar} shows the the stellar parameters of IL Cep and HD 216658 either generated in this work or compiled from the literature.
\begin{table}
\centering
\caption{The table provides the stellar parameters of IL Cep and HD 216658 used in this work}
\label{tab:stellar}
\begin{tabular}{ccccc}
\hline
Star & \begin{tabular}[c]{@{}c@{}}V \\ (mag)\end{tabular} & \begin{tabular}[c]{@{}c@{}}Distance\\ (pc)\end{tabular} & \begin{tabular}[c]{@{}c@{}}Spectral \\ type\end{tabular} & RUWE \\ \hline
IL Cep & 9.4 & 798$^{+8}_{-10}$ & B2-B3 & 1.0 \\
HD 216658 & 8.9 & 807$^{+24}_{-24}$ & B0.5V & 2.7 \\ \hline
\end{tabular}
\end{table}
In the new data from {\textit Gaia} EDR3, the star HD 216658 is at a distance of 807$^{+24}_{-24}$ pc, which is similar to the distance of IL Cep, .i.e 798$^{+8}_{-10}$ pc \citep{Bailer2020}. We saw from the astrometric analysis that HD 216658 is associated with IL Cep and other co-moving stars (\autoref{Fig2}). The star is excluded from the initial astrometric analysis due to its high RUWE parameter, which is 2.7 in the {\textit Gaia} EDR3. The value increased from the last release. Even though the RUWE parameter is high, we can indirectly confirm the association of HD 216658 with the IL Cep co-moving stars and it is part of the Cep OB3 association, which is reported to be at a distance of 800 pc \citep{Moreno1993A&A...273..619M, Pozzo2003MNRAS.341..805P}. The distances to the individual members estimated in the pre-{\textit Gaia} era were in the range of 500-1000 pc \citep{Crawford1970AJ.....75..952C}. We took the stars reported as part of the Cep OB3 association by \cite{Jordi1995A&AS..114..489J} and cross-matched them with {\textit Gaia} EDR3 and extracted the distances from \cite{Bailer2020}. The distance range of the sample was too large (180--9000 pc). Thus we took the stars with distances in the range of 500--1000 pc and found their transverse velocities. The median transverse velocity of the sample is 11.6 $\pm$ 0.6 km~s\textsuperscript{-1}. The transverse velocity of HD 216658 is 11.4 km~s\textsuperscript{-1} which is similar to the possible members of the Cep OB3 association.  

Similarly, the visual binary companion of HD 216658 is named in SIMBAD as HD 216658 B. This is at 6.6" separation from the primary and is a co-moving star candidate in our `level two' sample of IL Cep. Even though the proper motion values of HD 216658 and HD 216658B are shown to be distinct, the transverse velocity of the stars are 11.4 km~s\textsuperscript{-1} and 11.7 km~s\textsuperscript{-1} respectively, which is almost identical. Even with the {\textit Gaia} EDR3 data we cannot confirm that the visual binary of HD 216658 is bounded or not. But with the distance estimates and the transverse velocities, we can indicate that they are both part of the Cep OB3 cloud. This means the {\textit Gaia} EDR3 distance of HD 216658 is acceptable. Thus we consider the star as a co-moving candidate of Herbig Be star IL Cep.

\subsection{Two dynamically distinct population among co-moving stars}\label{sec:two_pop}

\begin{figure*}
   \centering
   \includegraphics[width=1.5\columnwidth]{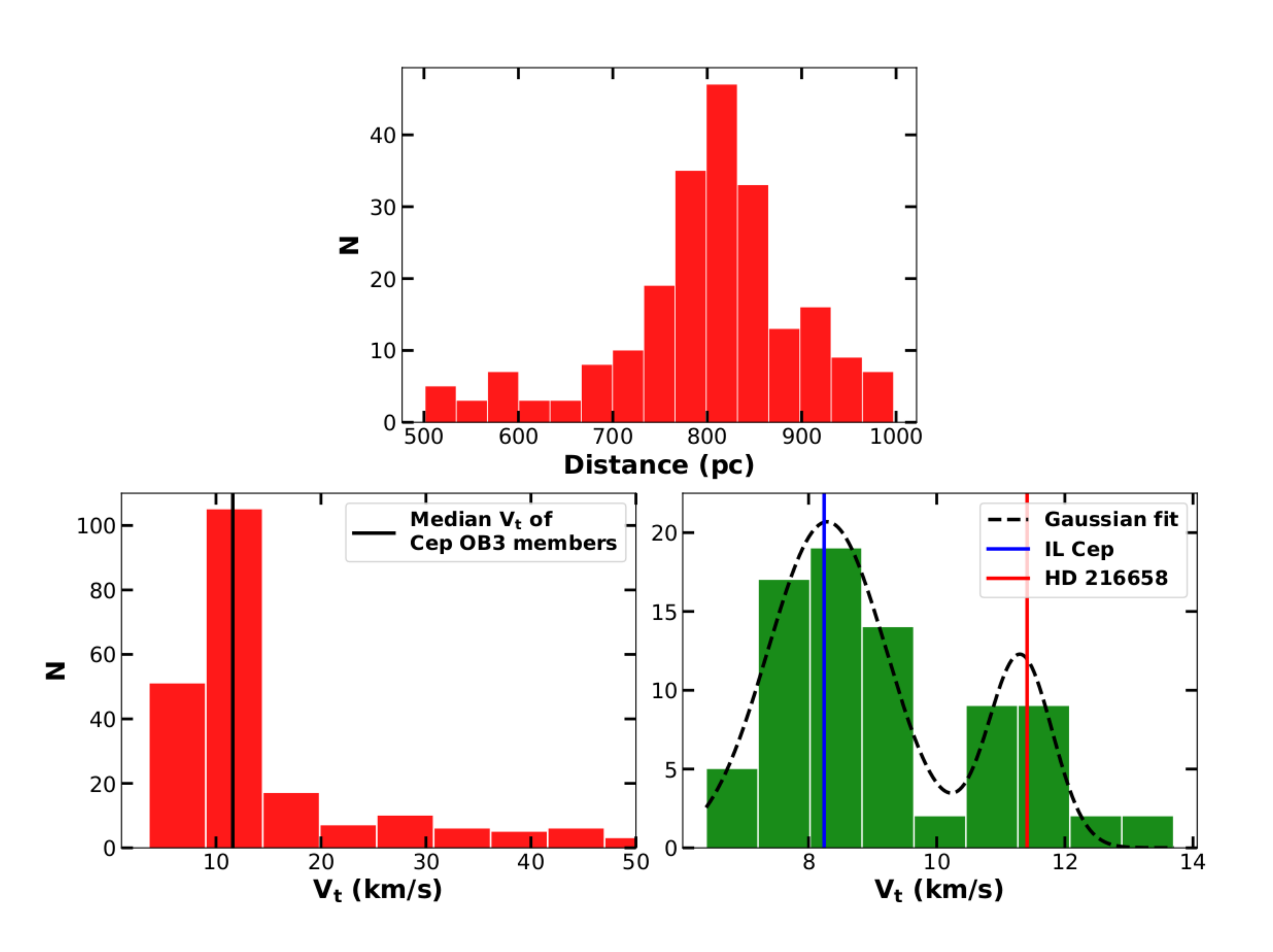}
   \caption{The figure shows three histogram distributions. (1) The distance distribution of Cep OB3 stars taken from \protect\cite{Jordi1995A&AS..114..489J} (top). (2) The transverse velocities of Cep OB3 stars (bottom left). (3) The transverse velocities of co-moving stars of IL Cep (bottom right). The distribution of transverse velocities of the co-moving stars is bimodal with two distinct populations associated with IL Cep and HD 216658. The transverse velocity of both stars is shown using red and blue vertical lines. The two-gaussian fit gave the mean transverse velocity of both populations as 8.3 km~s\textsuperscript{-1} and 11.4 km~s\textsuperscript{-1}. The distance distribution peaks at 800 pc, which is similar to the co-moving members of IL Cep. Also, the transverse velocity distribution of the sample has a median value of 11.6 $\pm$ 0.6 km~s\textsuperscript{-1}. This is similar to the population of HD 216658.}
   \label{Fig4}
   \end{figure*}

We estimated the transverse velocity of 79 co-moving stars of IL Cep including HD 216658 using {\textit Gaia} EDR3 astrometric data. The histogram distribution of the transverse velocities of the stars is shown in \autoref{Fig4} (bottom right). The distribution appears to be bimodal, which indicates the presence of two populations of stars among the highly constrained co-moving candidates. The transverse velocity of IL Cep and HD 216658 show that both stars associated with the same cloud, possess different transverse velocities. The bimodal histogram is fitted with a two-gaussian function and the mean velocities of the two populations are found to peak at 8.3 km~s\textsuperscript{-1} and 11.4 km~s\textsuperscript{-1}, respectively. This is consistent with the transverse velocity of IL Cep and HD 216658, which are 8.2 km~s\textsuperscript{-1} and 11.4 km~s\textsuperscript, respectively. The average uncertainty of the transverse velocity of stars in the analysis is 1.2 km~s\textsuperscript{-1}. Thus the two populations having a mean velocity difference of 3.2 km~s\textsuperscript{-1} should be real.  
   
The total stars are classified as two populations, dividing them at a velocity of 10.3 km~s\textsuperscript{-1} which is the intersection of both Gaussians. Population one is called the HD 216658 population with 24 stars and population two is called the IL Cep population with 56 stars. The IL Cep population constrained by astrometric parameters and transverse velocity is labeled as "IL Cep stellar group". Both populations are distinctly represented in the \autoref{Fig3}. The HD 216658 population does not show any preferential clustering around HD 216658 whereas the stars in the IL Cep population seem to be clustered around IL Cep. \autoref{Fig4} is additionally showing the histograms made using the distances and transverse velocities of Cep OB3 stars taken from \cite{Jordi1995A&AS..114..489J}. The Cep OB3 stars have a distance distribution peaking at 800 pc, which is at a similar distance as the co-moving stars of IL Cep. Also, the transverse velocity distribution of the sample has a median value of 11.6 $\pm$ 0.6 km~s\textsuperscript{-1}. This is similar to the population of HD 216658. This could indicate that the HD 216658 population is a part of the bigger parent cloud Cep OB3 and IL Cep is a more recent star formation inside the Cep OB3 cloud with a slightly lesser transverse velocity due to some pressure variations inside the parent cloud. From hereon the co-moving stars identified in this study will be treated as two distinct sub-populations associated with HD 216658 and IL Cep, respectively.  

\subsection{HD 216658 - Central exciting source}\label{sec:HD 216658_II}

The star HD 216658 is found to be astrometrically associated with the IL Cep co-moving stars. The star is at a similar distance as IL Cep. We also identified two populations of stars from the transverse velocity analysis of co-moving stars, each corresponding to IL Cep and HD 216658. Being the most massive star in the region, HD 216658 should be the predominant excitation source, which created the cavity. \cite{Deharveng2010A&A...523A...6D} noticed that the infrared bubble-like structures created by OB stars can be traced by the 8 $\mu m$ emission due to polycyclic aromatic hydrocarbon (PAH) molecules. Also, one can see 24 $\mu m$ emission in the central regions of the bubble, which is due to the emission from the dust grains. \cite{Deharveng2010A&A...523A...6D}, using \textit{Spitzer} 8 $\mu m$ (IRAC 4) and 24 $\mu m$ (MIPS 1) images, pointed out that the bubbles/cavity can be traced by the 8 $\mu m$ images which shows PAH dominated emissions and the central source is associated with 24 $\mu m$ warm dust emission around it. 

\begin{figure}
   \centering
   \includegraphics[width=1\columnwidth]{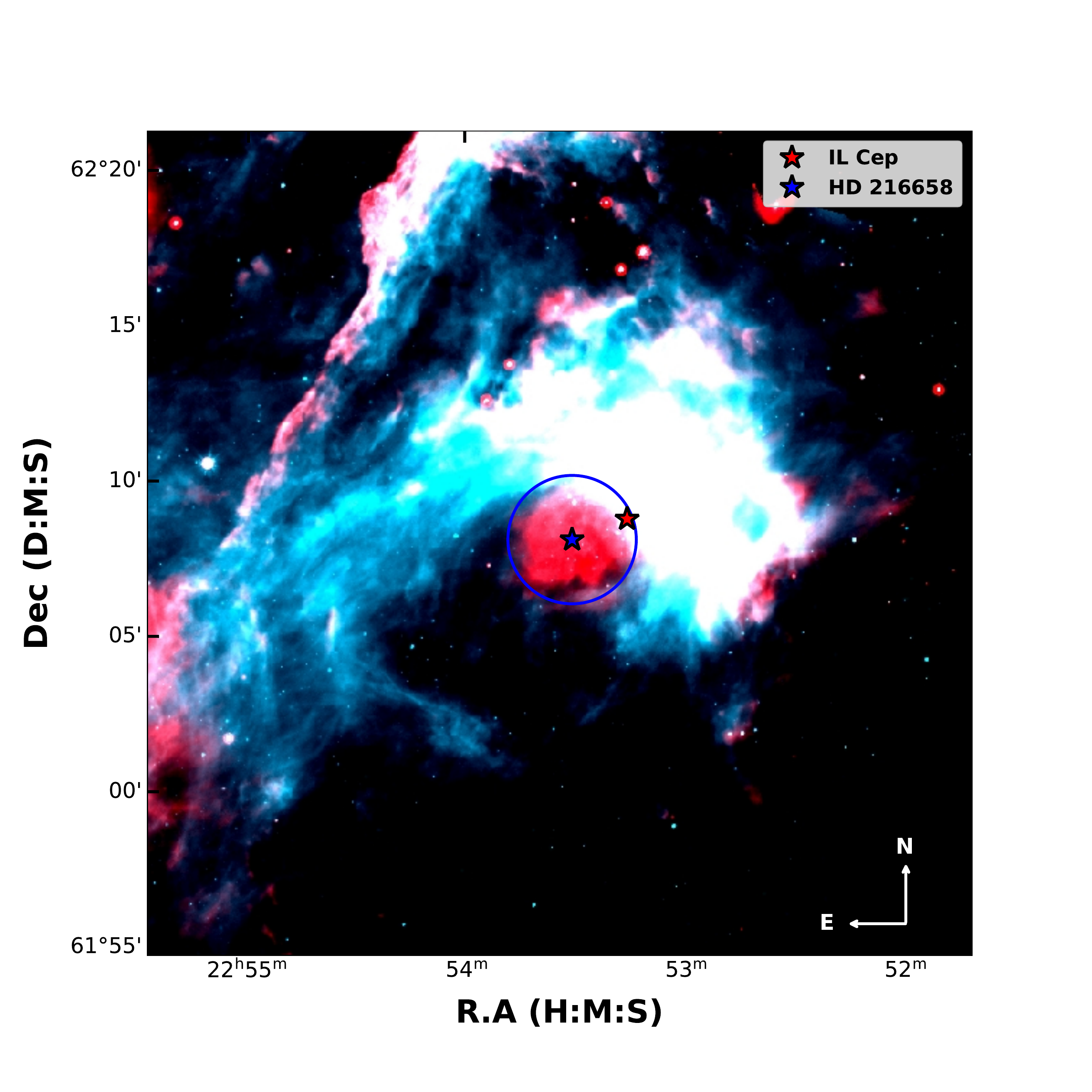}
   \caption{ The \textit{Spitzer} color composite image of the region around HD 216658 created using 8 $\mu m$ (IRAC 4) emission in turquoise and 24 $\mu m$ (MIPS 1) in red. The Strömgren radius of HD 216658 is shown as a blue circle around HD 216658. The spherical distribution of 24 $\mu m$ emission inside the Strömgren radius indicates that the cavity is formed by HD 216658.}
   \label{Fig5}
   \end{figure}

\autoref{Fig5} shows the \textit{Spitzer} color composite image of the region around IL Cep, similar to the images generated by \cite{Deharveng2010A&A...523A...6D}. The figure shows both 8 $\mu m$ and 24 $\mu m$ emissions. We calculated the Strömgren radius for HD 216658 assuming the spectral type of the star as B0V \citep{Morgan1953ApJ...118..318M, Garrison1970AJ.....75.1001G}. The radius is estimated to be 0.5 pc (\textasciitilde 124 arcsec), which is illustrated in the \autoref{Fig5}. We can see a spherical distribution of 24 $\mu m$ emission inside the Strömgren radius. The region around HD 216658 could be an ionised region completely devoid of 8 $\mu m$ PAH emission. This indicates that the predominant excitation source of the region is HD 216658, which may also be responsible for the creation of the cavity.

The astrometric solution of HD 216658 appears to be bad (RUWE = 2.7). But the star's geometrical position, association with a transverse velocity population which is identical to Cep OB3 stars, and the 24 $\mu m$ emission in its Strömgren radius show that HD 216658 is associated with the cavity and the Herbig Be star IL Cep.  

\subsection{Coevality of co-moving stars from {\textit Gaia} color-magnitude diagram }\label{sec:CMD}
 
 \begin{figure*}
   \centering
   \includegraphics[width=1\columnwidth]{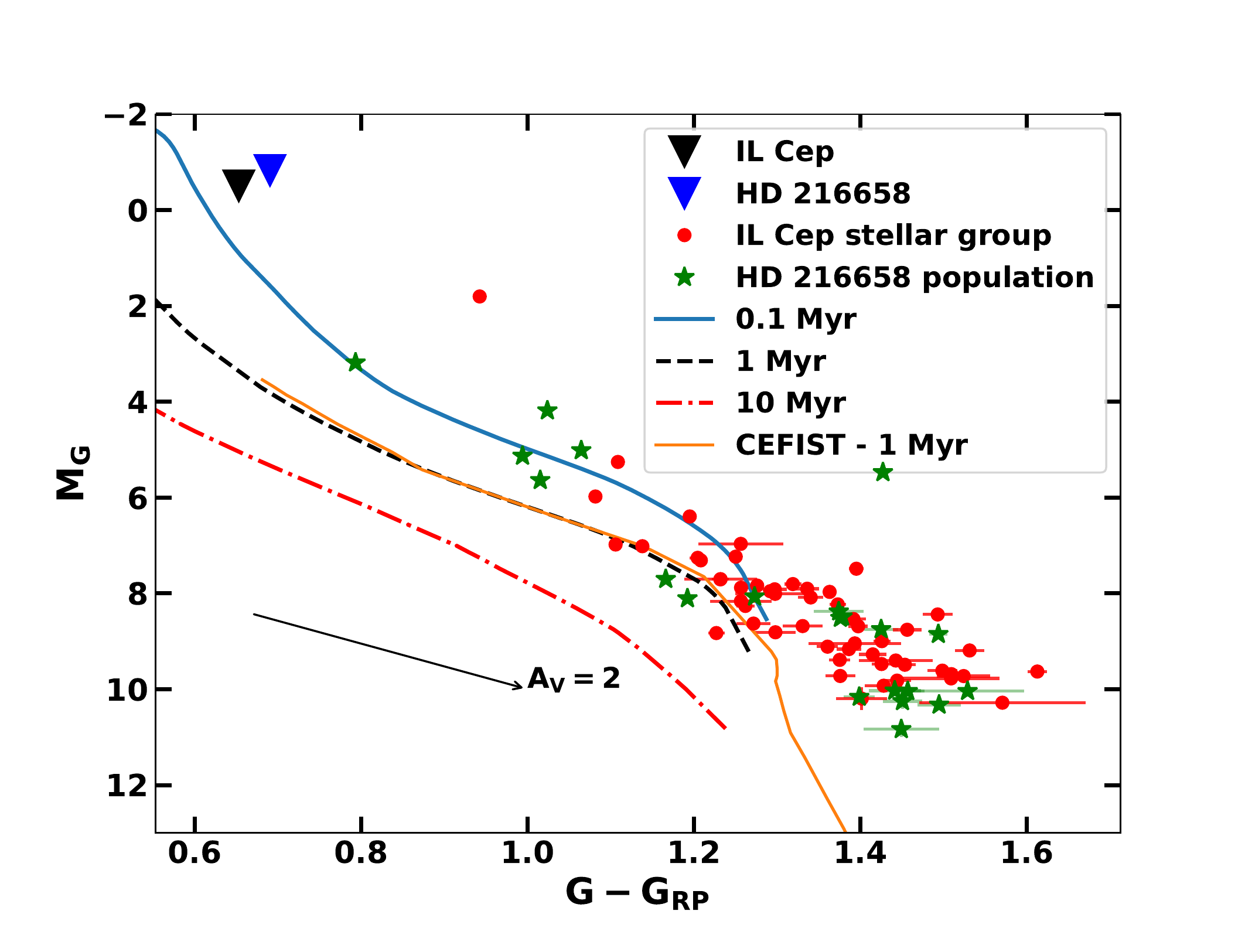}
   \includegraphics[width=1\columnwidth]{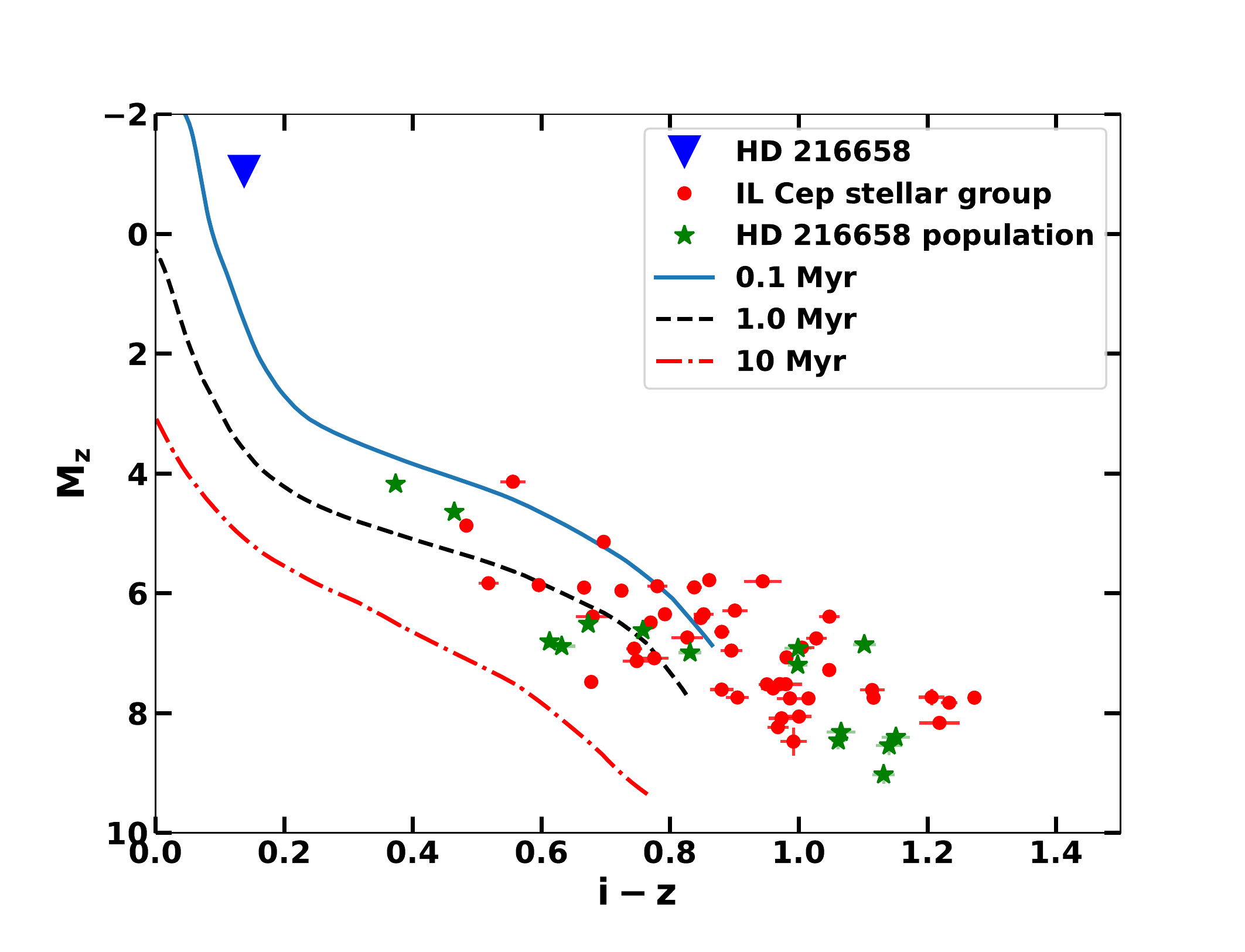}
\caption{The {\textit Gaia} EDR3 (left) and Pan-STARRS (right) CMD of the co-moving stars are illustrated. Most of the co-moving stars are identified to be coeval to IL Cep and are all indeed PMS stars in both CMD. IL Cep and HD 216658 are shown as black and blue star symbols respectively in the {\textit Gaia} CMD. Only HD 216658 is shown in Pan-STARRS CMD. }
   \label{Fig6}
\end{figure*} 

We identified 79 co-moving sources associated with IL Cep from the astrometric analysis of {\textit Gaia} EDR3 data. The clustering of these stars around IL Cep suggests that they may have formed at the same time as IL Cep formed. Also, we found that there are two distinct transverse velocity populations in the co-moving stars. \cite{Arun2019} estimated the age of Herbig Be star IL Cep as 0.11 $\pm$ 0.1 Myr from the isochrone fitting on the Gaia color-magnitude diagram. We used the {\textit Gaia} photometric magnitudes \citep{Riello2020EDR3_CMD} and distances \citep{Bailer2020} for plotting the {\textit Gaia} color-magnitude diagram (CMD) (G-G\textsubscript{RP} vs M\textsubscript{G}) of the co-moving sources \citep{ Kiman2019AJ....157..231K}. \autoref{Fig6} shows the {\textit Gaia} CMD of newly identified co-moving stars along with IL Cep and HD 216658. We have represented the subpopulations associated with IL Cep and HD216658 as separate symbols in the CMD to see whether they show any age difference. Modules for Experiments in Stellar Astrophysics (MESA) isochrones and evolutionary tracks (\href{http://waps.cfa.harvard.edu/MIST/}{MIST})\footnote{http://waps.cfa.harvard.edu/MIST} of ages 0.1, 1, and 10 Myr \citep{choi2016, Dotter2016} along with low mass isochrone of 1 Myr from the CIFIST
2011\_2015\footnote{https://phoenix.ens-lyon.fr/Grids/BT-Settl/CIFIST2011\_2015/} is plotted in the CMD. The {\textit Gaia} CMD is not corrected for extinction. This will not affect the age estimates considerably since the interstellar extinction vector (A\textsubscript{V}) is approximately parallel to the isochrones, as shown in the \autoref{Fig6}. The CMD shows that majority of the co-moving stars are positioned above the 0.1 Myr isochrone. From the location of the co-moving stars in the CMD, it is identified that they are coeval to IL Cep. Most of the co-moving stars are positioned at regions slightly younger than IL Cep ( > 0.1 Myr). We may not be able to accurately estimate the ages of these stars. 

Another optical CMD is created using Pan-STARRS DR1 \citep{Chambers2016} photometric data to compliment the {\textit Gaia} CMD. The coordinates of the co-moving stars are cross-matched with Pan-STARRS database with a 3" radius and got 78 matches. Avoiding stars without i and z magnitudes and also the stars with i and z magnitude uncertainty less than 0.02 mag are retained. The Pan-STARRS $M\textsubscript{z}$ vs $(i-z)$ color-magnitude diagram (CMD) of 62 stars satisfying the criteria shown in \autoref{Fig6} (right). The MIST isochrones of ages 0.1, 1, and 10 Myr are also shown. The Pan-STARRS CMD also shows that the co-moving stars are a younger population ( $\sim$ 0.1 Myr). The i magnitude is not available for IL Cep. So only HD 216658 is also shown in the Pan-STARRS CMD.   

 The age of IL Cep estimated from this study using {\textit Gaia} EDR3 matches with the previous estimates by \cite{Arun2019} using {\textit Gaia} DR2 data. Also, from the CMD analysis, we found that the co-moving stars are coeval to IL Cep. The CMD analysis confirms that co-moving stars around IL Cep are PMS stars and most of them are formed at similar timescales as IL Cep.  Interestingly, the brightest star in the region, HD 216658 is also occupying a similar position as IL Cep in the CMD. Also, from the present analysis, we do not find any distinction in ages between the sub-populations associated with HD 216658 and IL Cep.

\subsection{Identification of circumstellar disk candidates}\label{sec:ccd}

\begin{figure*}
   \centering
   \includegraphics[width=1\columnwidth]{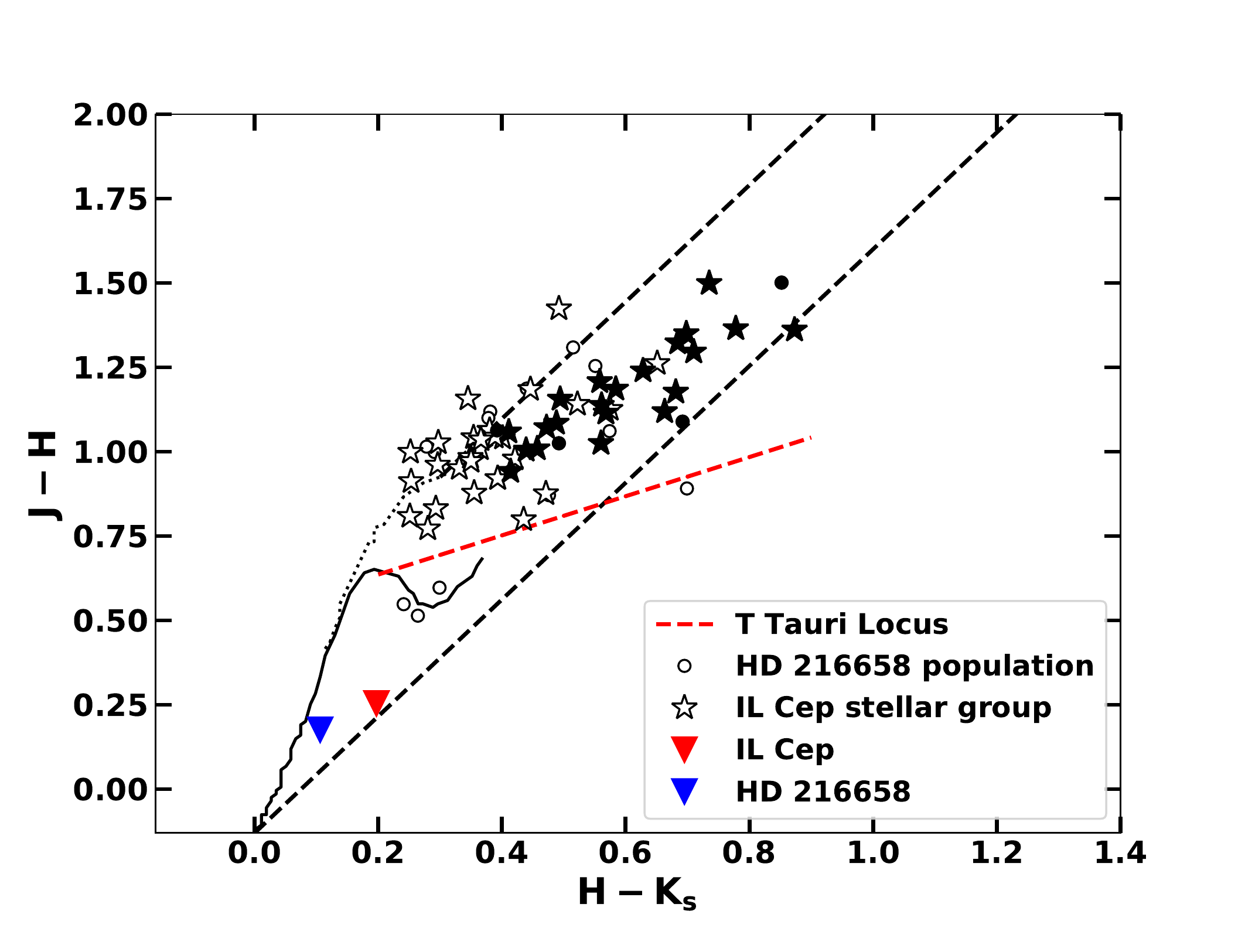}
   \includegraphics[width=1\columnwidth]{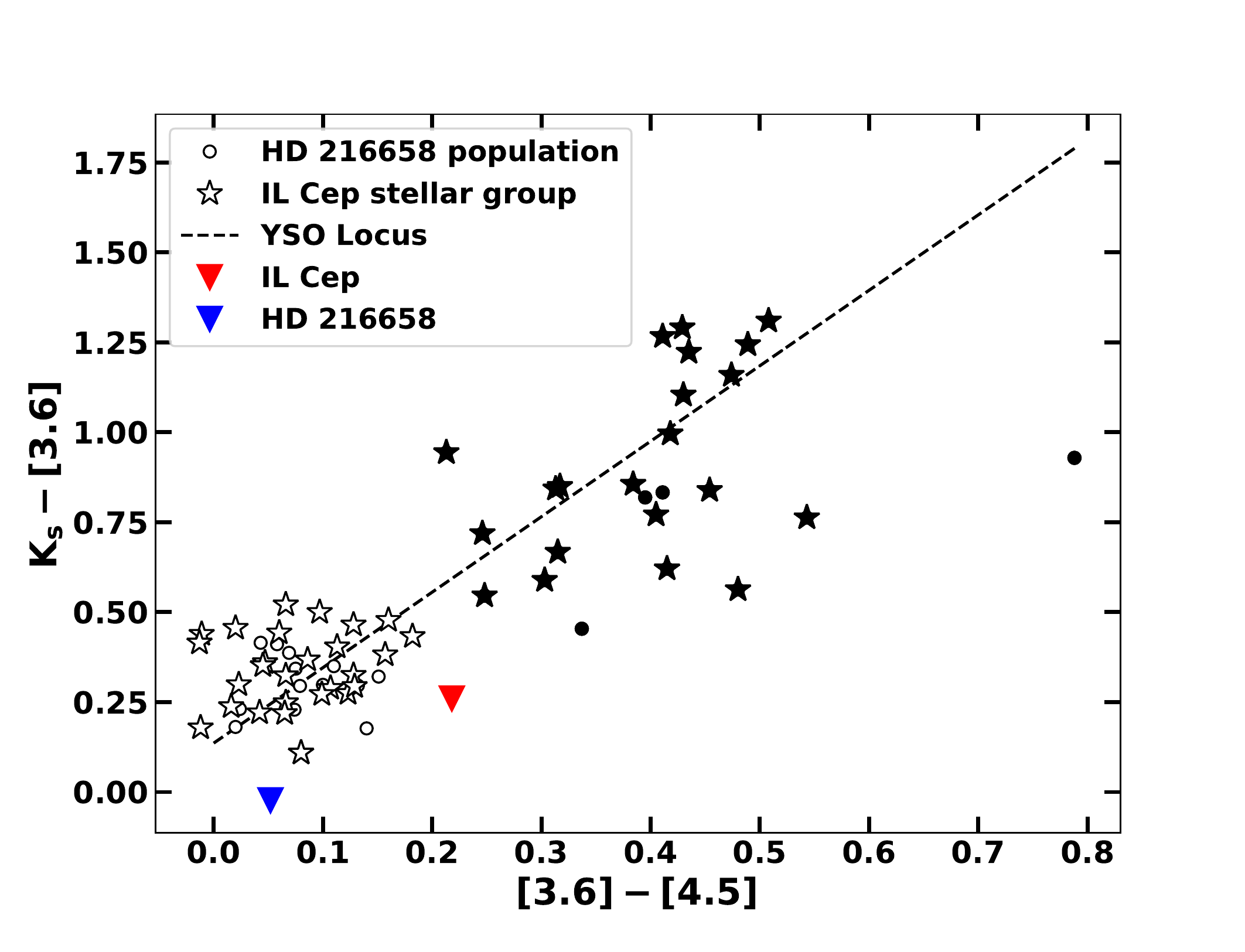}
\caption{Figures show 2MASS and IR CCDm of co-moving stars of IL Cep. The HD 216658 population is shown as circles and IL Cep stellar group is marked as star symbols. The Class II sources among both populations are shown with filled black colors. The Class III sources are represented by non filled symbols. IL Cep and HD 216685 are also shown in both figures.}
   \label{Fig7}
\end{figure*} 
 From the {\textit Gaia} CMD analysis we found that all the co-moving stars are coeval with IL Cep and they are PMS stars in the age range 0.1 -- 1 Myr. The stars with ages of a few Myr should be actively accreting from its circumstellar disk \citep{Furlan2011ApJS..195....3F, Semenov2020ENews..51a..29S}. The accreting young stars will show infrared excess in their spectral energy distribution due to the thermal re-radiation from the dust in the circumstellar disk \citep{Hillenbrand1992, Malfait1998}. \cite{Saha2020} found that 80\% of the co-moving stars around Herbig Be star HD 200775 are Class III objects. Thus it is important to identify the evolutionary phase of the co-moving stars of IL Cep. With adequate infrared data, we can classify the young co-moving stars around IL Cep as embedded protostars (Class I), PMS stars with a circumstellar disk (Class II), and those that have already dissipated their accretion disks (Class III objects). 
 
 We extracted the 2MASS \citep{Skrutskie2006} and \textit{Spitzer} Glimpse 360 \citep{Whitney2011AAS...21724116W} data of the co-moving stars of IL Cep. Glimpse 360 data contain IRAC [3.6] and IRAC [4.5] magnitudes but it does not include IRAC [5.8] and IRAC [8] magnitudes. The stars with 2MASS magnitudes with the quality flag `AAA' and Glimpse 360 IRAC [3.6] and [4.5] magnitudes with uncertainty < 0.1 mag are used for the identification of IR excess candidates. We selected 70 stars that satisfy the magnitude quality criteria. For the identification of IR excess candidates, we adopted the method developed by \cite{Gutermuth2008ApJ...674..336G} using 2MASS JHK magnitudes and IRAC [3.6], [4.5] magnitudes. The method uses the T Tauri locus defined by \cite{Meyer1997} to find the intrinsic colors and the color constraints, which is explained in detail by \cite{Gutermuth2008ApJ...674..336G}. Using this method, we identified 25 stars as Class II objects and the remaining stars as Class III objects. The 2MASS H-K\textsubscript{s} vs J-H color-color diagram (CCDm) and the IR CCDm using ([3.6] - [4.5]) vs (K\textsubscript{s} - [4.5]) are shown in \autoref{Fig7}. The stars of each sub-population among the co-moving stars are illustrated in the figure separately. The method adopted from \cite{Gutermuth2008ApJ...674..336G} provided the A\textsubscript{V} value relative to the T Tauri locus of all the Class II objects. The average extinction of the region is calculated as A\textsubscript{V} = 3.7 mag by taking the mean extinction of Class II sources. Also, we observe that 65\% of all the co-moving stars are diskless sources (Class III). This observation is consistent with the findings of \cite{Saha2020} in the case of Herbig Be star HD 200775, where almost 80 \% of the co-moving sources identified are Class III. 

\begin{figure}
   \centering
   \includegraphics[width=1\columnwidth]{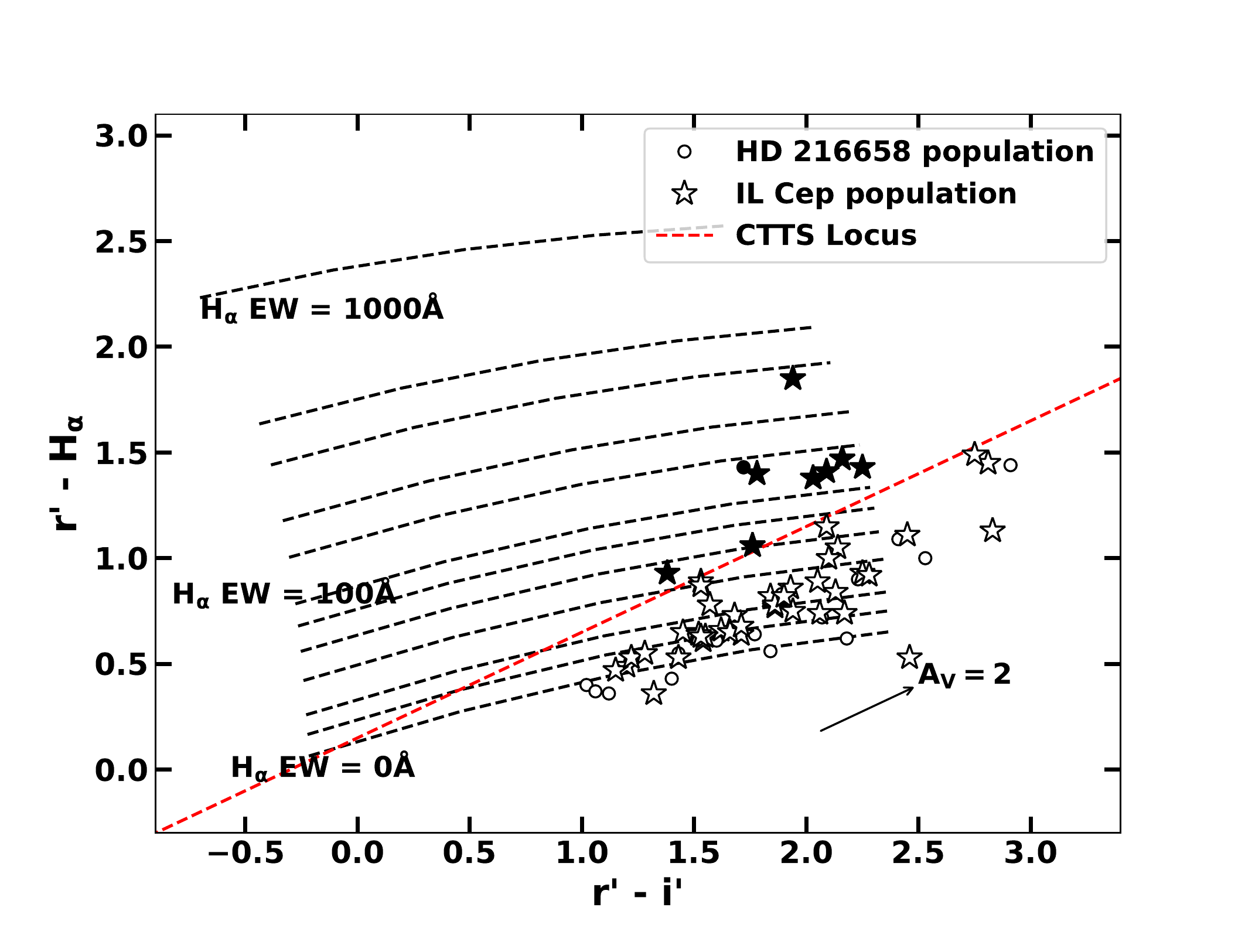}
   \caption{The figure shows the IPHAS CCDm with the synthetic H$\alpha$ EW grid. The color-coding of the sources is similar to \autoref{Fig7}. The dark-filled symbols are H$\alpha$ sources identified in the study. }
   \label{Fig8}
   \end{figure}
\subsection{Identification of H$\alpha$ emission sources}\label{sec:iphas}   
The PMS stars are generally classified as T Tauri \citep{Joy1945ApJ...102..168J} and HAeBe stars \citep{Herbig1960}. The PMS stars have the common property of showing emission lines in their spectra \citep{Hillenbrand1992}. We classified the co-moving stars into Class II and Class III objects in the previous section. In this section, we identify H$\alpha$ emitting stars among the co-moving sources around IL Cep. We did not avoid Class III objects for the analysis as there can be weak line T Tauri stars among the photometrically classified Class III objects. Also, \cite{Saha2020} identified Class III objects showing weak H$\alpha$ emission from spectroscopy. We used the photometry from IPHAS \citep{Drew2005MNRAS.362..753D, Barentsen2011MNRAS.415..103B} which provides narrowband H$\alpha$ photometry along with Sloan r' and i' magnitudes. We use IPHAS colors to construct CCDm for the identification of emission-line sources. 

The IPHAS DR2 data for the stars are taken from the Vizier catalog services. We used a 3" search radius for all the co-moving stars associated with IL Cep for finding the IPHAS counterparts. If two detections are reported for a star, we took the closest one to the given coordinates. The IPHAS CCDm (r'-i') vs (r'-H$\alpha$) is shown in \autoref{Fig8}. The synthetic grid for various H$\alpha$ emission equivalent width values are shown in Figure 8, which is adopted from \cite{Drew2005MNRAS.362..753D}. We used the classical T Tauri star (CTTS) locus to identify intense H$\alpha$ emission sources \citep{Damiani2017A&A...602A.115D, Damiani2018A&A...615A.148D}. There are 9 sources above the CTTS locus, and they can be considered as confirmed H$\alpha$ emission sources. All the 9 emission sources are classified as Class II sources in the infrared CCDm analysis in the previous section. All Class III sources and some Class II sources are below the CTTS locus.  We may have missed some of the less intense H$\alpha$ emission sources in the analysis. A thorough spectroscopic observation is required to identify less-intense H$\alpha$ sources.

\begin{figure*}
   \centering
   \includegraphics[width=1\columnwidth]{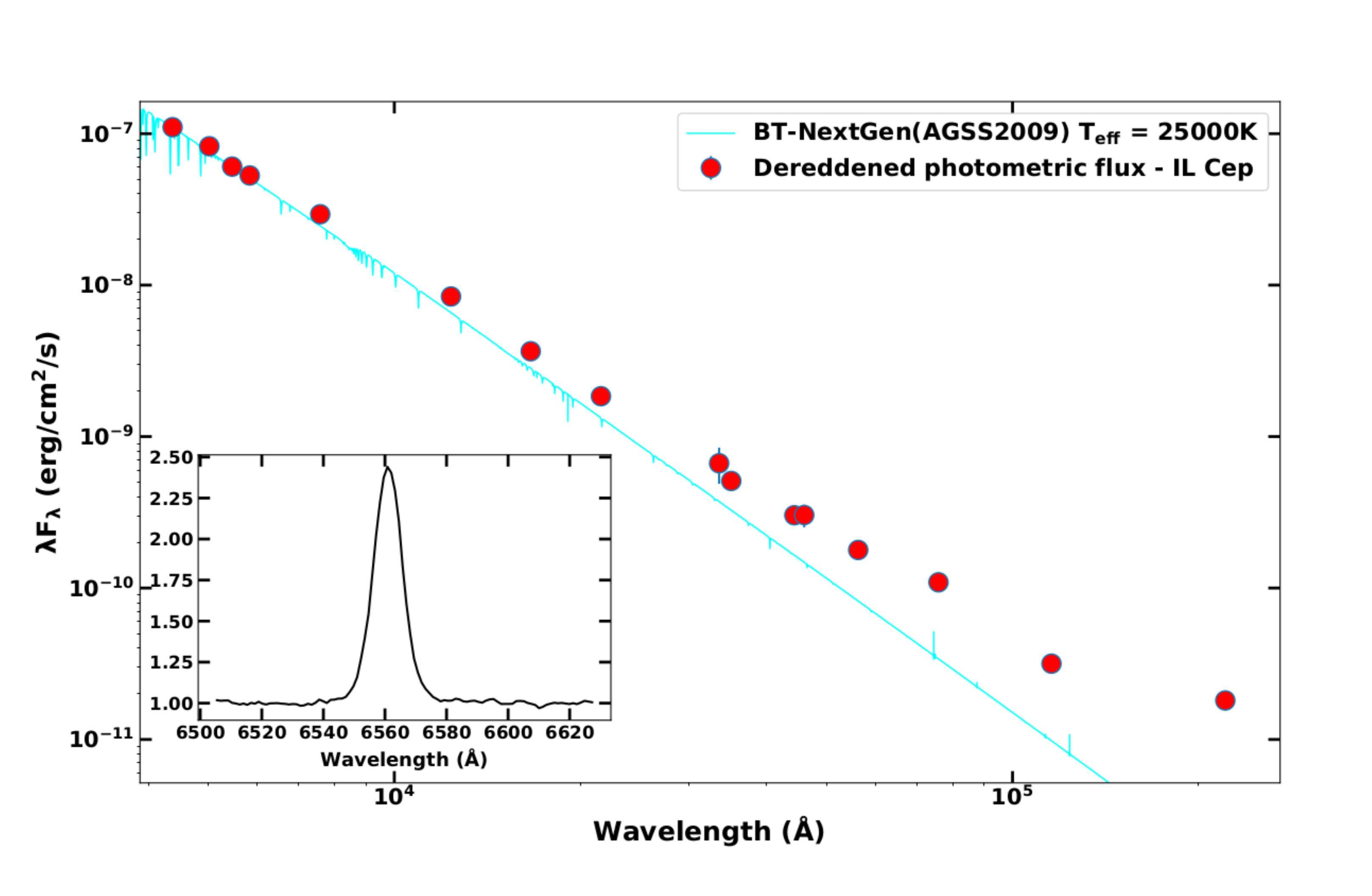}
   \includegraphics[width=1\columnwidth]{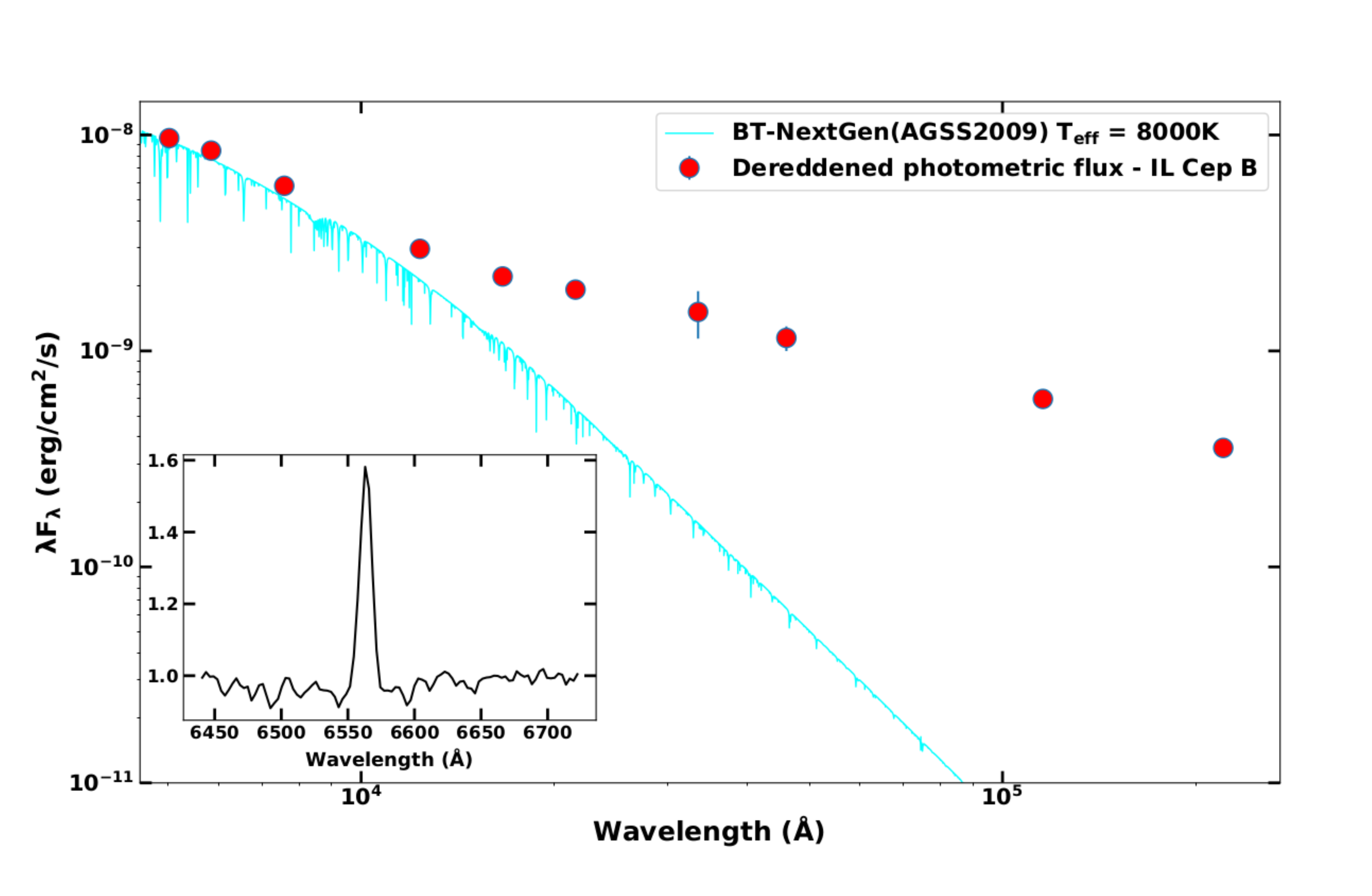}
\caption{Figures show the best fit SEDs of IL Cep and IL Cep B. The BT-Next Gen (AGSS2009) synthetic spectra of fitted T\textsubscript{eff} are shown. The H$\alpha$ emission profiles of both stars are shown on the bottom left corners of each SED.}
   \label{Fig9}
\end{figure*}
\subsection{Spectroscopy and SED of bright stars}\label{sec:sed}
\subsubsection{IL Cep multiple system}\label{sec:sed_1}

IL Cep is reported to be an unresolved binary star by \cite{Wheel2010} and \cite{Ismailov2016A}. In this section, the star mentioned as IL Cep B (HD 216629B) is the visual binary companion of IL Cep \citep{MelNikov1996ARep...40..350M}. For brevity, we will be mentioning IL Cep A as IL Cep in this work. The companion star IL Cep B is at a separation of 7" from the primary \citep{EDR3_I}. The binary companion is also among the co-moving stars of IL Cep identified using {\textit Gaia} EDR3 astrometric analysis (see Sect.\ref{sec:astrometry}). The star is the second brightest star in the co-moving group, the brightest being IL Cep itself. Both the stars being luminous and close to each other, IPHAS magnitudes are not listed in the archive and are not included in the emission line star identification analysis. We observed the optical spectra of IL Cep using the Himalayan Faint Object Spectrograph Camera (HFOSC) mounted on the 2-m Himalayan Chandra Telescope (HCT). The spectrum of IL Cep is reported in the study of \cite{Mathew2018}, which discusses primarily about the O{\sc i} lines in HAeBe stars rather than elaborating the spectral details of IL Cep. Further, we observed the spectrum of IL Cep B during our recent observation run using the OMR spectrograph (details are given in Sect. \ref{sect:data}). It may be noted that although the spectra of both stars are taken with different spectrographs, the resolution is similar ($\sim$ 8 \AA).

The H$\alpha$ emission profiles of both stars are shown in \autoref{Fig9}. The H$\alpha$ emission is reported in IL Cep in various studies and the present data confirms the emission. Interestingly, we found that IL Cep B also shows emission in H$\alpha$. The measured H$\alpha$ equivalent width (EW) of IL Cep and IL Cep B are  -16 \AA~ and -7 \AA~, respectively. The spectrum of IL Cep B has low SNR, making it difficult to estimate a proper spectral type. We constructed the SED of IL Cep and IL Cep B using the available magnitudes in optical and IR passbands. The magnitudes are extinction corrected with an A\textsubscript{V} of 3.12 mag, taken from \citep{Vioque2018A&A...620A.128V}. The IL Cep magnitudes in optical passbands are fitted (with chi-squared minimization) with a theoretical stellar atmosphere of effective temperature (T\textsubscript{eff}) 25000 K, at solar metallicity and for a surface gravity ($log g$) of 4.5. Similarly, for IL Cep B, the U, B, V magnitudes are fitted well with the stellar atmosphere corresponding to solar metallicity for T\textsubscript{eff} = 8000 K and $log g$~ = 4.5. The spectral type is further validated from the temperature using the look-up table in \cite{mamajeck2013}. The spectral type of IL Cep is found to be B1.5, in agreement with the studies of \cite{Morgan1953ApJ...118..318M} and \cite{Garrison1970AJ.....75.1001G}. IL Cep B is found to be an HAe star of spectral type A6. The SEDs of both stars show IR excess, with IL Cep B showing higher flux excess than IL Cep. The spectral index (Lada index: \citealp{LAda1987,Green1994}) for both stars are estimated using 2MASS K\textsubscript{s} and WISE W2 magnitudes ($n_{2-4.6}$; \citealp{Arun2019,Anusha2020MNRAS.tmp.3745A}). The $n_{2-4.6}$ for IL Cep and IL Cep B are -2.4 and -0.6 respectively. 

From the above discussion, we found that both the stars in the IL Cep binary system show H$\alpha$ emission and IR excess, which are characteristic features of HAeBe stars. This classifies IL Cep to the rare class of binaries where one component is a Herbig Be star and the other is a Herbig Ae star

\subsubsection{HD 216658}\label{sec:sed_2}

\begin{figure*}
   \centering
   \includegraphics[width=1\columnwidth]{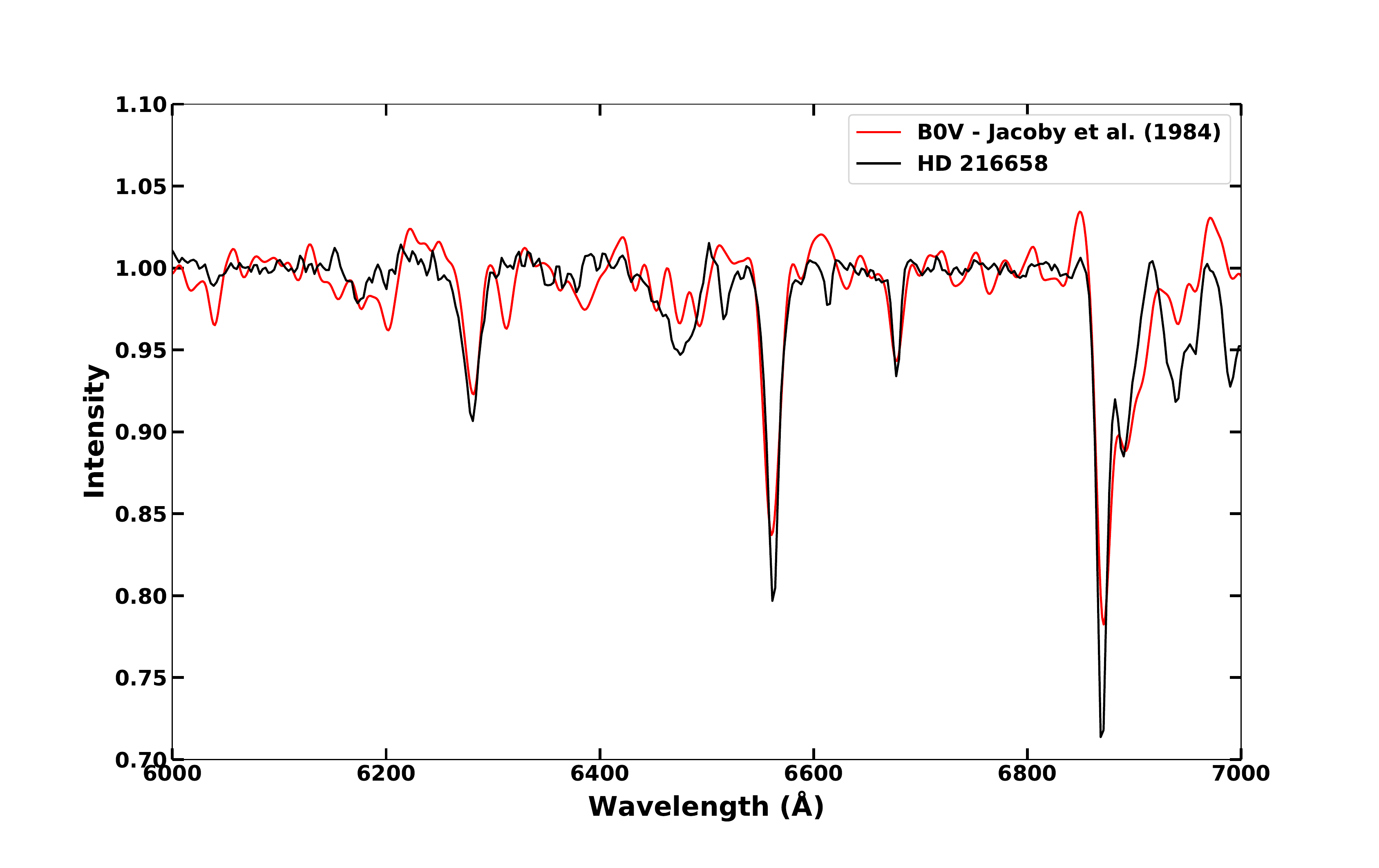}
   \includegraphics[width=1\columnwidth]{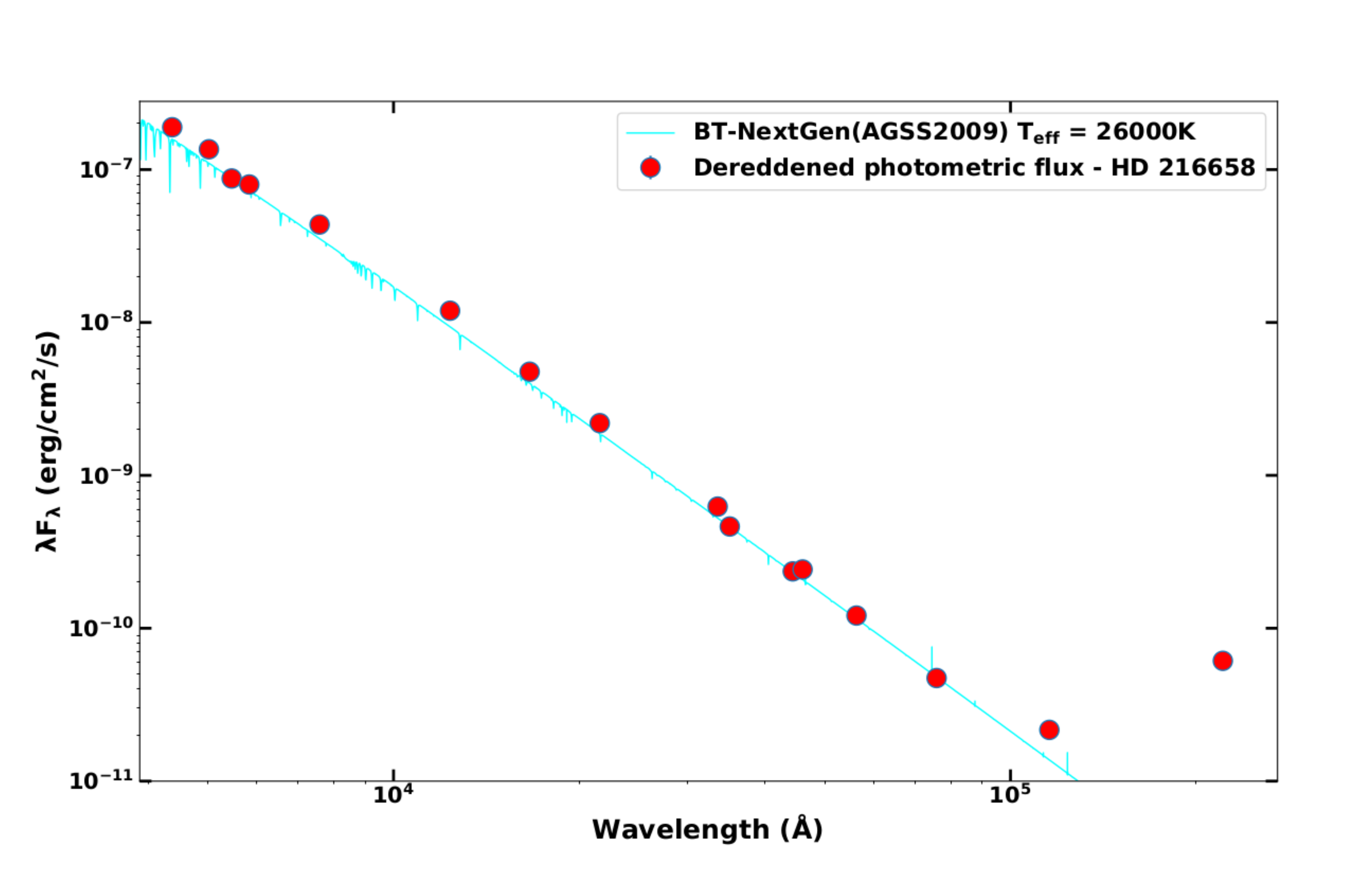}
\caption{Figures show the optical spectrum (left panel) and SED (right panel) of HD 216658. The spectral type of the star is estimated to be B0V using \protect\cite{Jacoby1984ApJS...56..257J}. The library spectra with spectral type B0V is also shown in the figure. The SED shows HD 216658 not having any near-IR excess.}
   \label{Fig10}
\end{figure*}
The star HD 216658 is positioned at a similar location as IL Cep in the {\textit Gaia} CMD. The star is identified to be the predominant excitation source in the region. The optical spectrum of HD216658 in the wavelength range 6000 -- 7000 \AA~is shown in \autoref{Fig10}. The star does not show any emission lines in the observed spectrum. The spectral type of HD 216658 is determined as B0V by comparing the absorption strength of He{\sc i} 6678 \AA~with that of main sequence stars from the stellar library of \cite{Jacoby1984ApJS...56..257J}. This estimate is consistent with the previous spectral types reported in the literature, which is B0V - B0.5V \citep{Morgan1953ApJ...118..318M, Garrison1970AJ.....75.1001G}. Before performing the comparison, we normalized the spectra of HD 216658 and took the templates to a common resolution as the VBT spectra.
The uncertainty in our spectral classification is found to be of two spectral subclasses. The age of IL Cep and other co-moving stars is approximately 0.1 Myr. The spectral type being B0V, we estimated the PMS timescale of HD 216658 to be 0.03 Myr. If we assume that IL Cep stellar group is formed almost 0.1 Myr ago, HD 216658 completed its PMS evolutionary stage in at least 0.05 Myr and the inner disk must have got cleared in this timescale. This is consistent with the emission-less spectra for HD 216658. The SED fitted on the photometric data using the BT-NextGen model for HD 216658 is also shown in \autoref{Fig10}. The star does not show IR excess till the WISE W3 band and a rise in SED is seen in the W4 band. The excess in W4 may be due to the IR emission from dust around HD 216658 (Sect. \ref{sec:HD 216658_II}). However, it may be noted that the W4 magnitude is flagged as "d" in the WISE database, which means there is a possible diffraction spike on the observed image \citep{Cutri2013}. Hence, we cannot consider the excess in the W4 band to be real.

The star HD 216658 is found to be a member of the co-moving stars associated with IL Cep. The star is the most massive in the region and appears to be more evolved when compared to other co-moving stars, including IL Cep, and appears to have completed its PMS phase. Thus the star being more evolved and at a favorable position geometrically, it is safe to assume that HD 216658 is the initial trigger for the formation of the cavity in the region. IL Cep may have formed in the rims of the expanding cavity, along with the low mass co-moving stars.

\subsection{Molecular clumps around IL Cep}\label{sec:clumps}

The astrometric analysis has provided 78 co-moving stars associated with IL Cep, out of which 26 have been identified as IR excess candidates. The region appears to be having stars of age $\sim$ 0.1 Myr. Also, previous studies have shown that there are pre-stellar clumps on the edges of the expanding shell-like structure called the "cavity" \citep{Zhang2016MNRAS.458.4222Z}. This indicates that the region is very young and is still forming stars. \cite{Zhang2016MNRAS.458.4222Z} only studied the northern part of the cavity and identified 6 molecular clumps using \textsuperscript{13}CO(J = 1-0) velocity-integrated intensity contours. To study all the regions around the cavity, especially the southern region, we use the H\textsubscript{2} column density map generated using the Herschel maps in four bands (160 to 500 $\mu m$) \citep{Andre2010A&A...518L.102A, Marsh2015MNRAS.454.4282M}. We extracted the point process mapping (PPMAP)\footnote{http://www.astro.cardiff.ac.uk/research/ViaLactea/PPMAP\_Results/} data of 25$'$ x 25$'$ region around IL Cep, which covers the molecular cavity defined by \cite{Zhang2016MNRAS.458.4222Z}. The PPMAP has an angular resolution of 12" and a pixel scale of 6". To identify the clump-like structures in the region we used the dendrogram algorithm \citep{Goodman2009Natur.457...63G}. The dendrogram represents the hierarchy of the structures in any data. In our case, it is the H\textsubscript{2} column density map. The python version of the algorithm \textit{astrodendro} is implemented on the PPMAP to identify the structures on or around the cavity. 

\begin{figure}
   \centering
   \includegraphics[width=1\columnwidth]{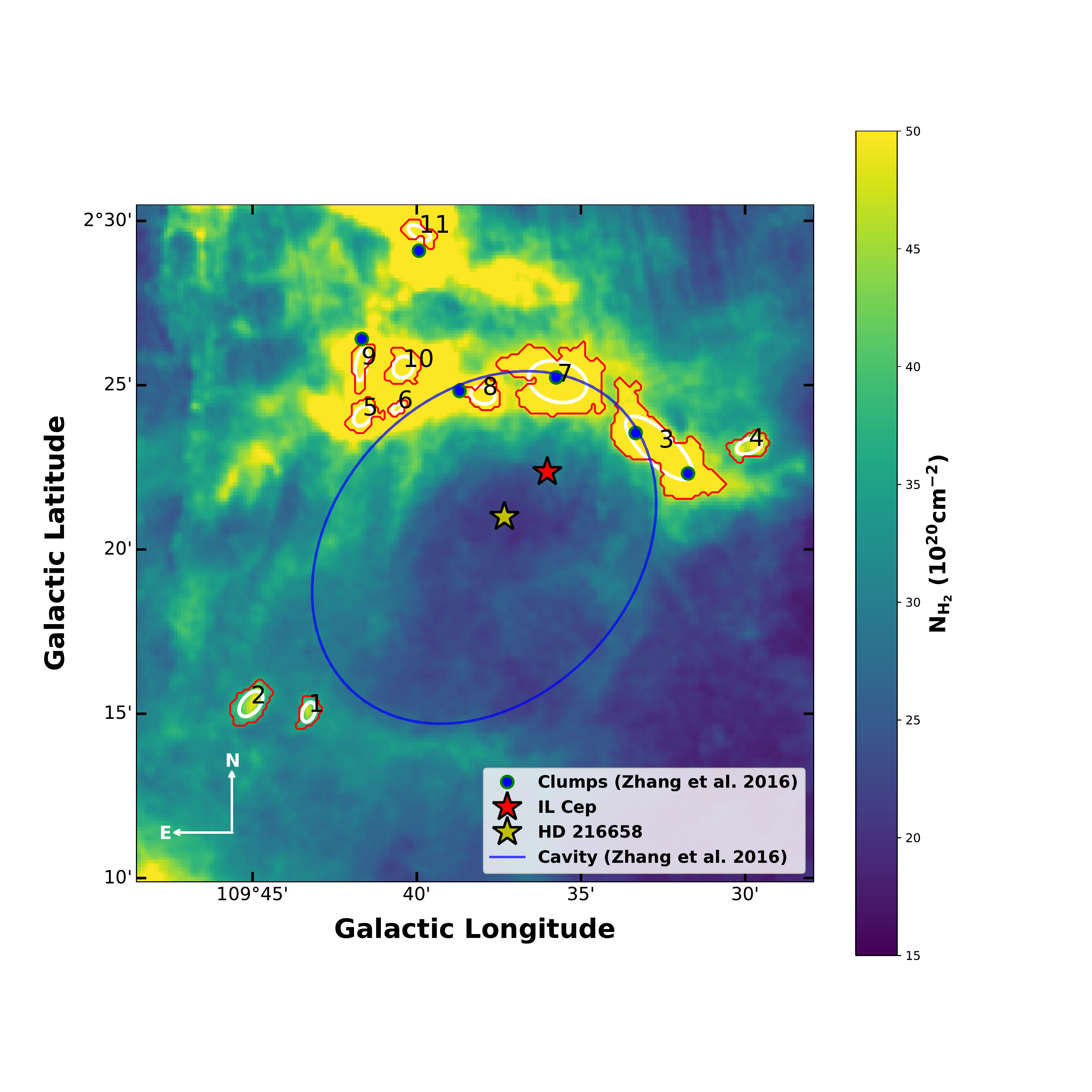}
\caption{Figure illustrates the Herschel column density map taken from PPMAP. Dendrogram analysis identified 11 molecular clump-like structures on the expanding "cavity" around IL Cep. The structures on the northwest side are in agreement with the clumps identified by \protect\cite{Zhang2016MNRAS.458.4222Z}. Also, we have identified two additional clumps on the south-eastern region of IL Cep.}
   \label{Fig11}
\end{figure}

 The \textit{astrodendro} algorithm requires three input parameters for the implementation of the structure identification routine. The parameters are the lowest background (\texttt{min\_value}), the lowest height of a structure (\texttt{min\_delta}), and the minimum pixel size of a structure (\texttt{min\_npix}). The \texttt{min\_value} and \texttt{min\_DELTA} are taken as 3$\sigma$ and 1$\sigma$, which is 35.7 $\times$ 10\textsuperscript{20} cm\textsuperscript{-2} and 11.9 $\times$ 10\textsuperscript{20} cm\textsuperscript{-2} respectively \citep{Walker2021arXiv210203560W}. In order to identify a structure, we set a threshold of 5 pixels, in accordance with the PPMAP resolution \citep{Watkins2019A&A...628A..21W}. Avoiding all the structures on the sides of the frame, we identified 11 structures on and around the cavity. The column density map and the identified clumps are shown in \autoref{Fig11}. Most of the northern structures we identified are consistent with the clumps identified by \cite{Zhang2016MNRAS.458.4222Z}. Besides, we identified two clumps on the southern side of the cavity that was not identified in the previous studies. We estimated the total mass of each structure using the radius and N(H\textsubscript{2}) values derived from the dendrogram algorithm. The coordinates, radius, and mass of the structures are listed in \autoref{tab:2}. This implies that the ionization front created by the central star HD 216658 has compressed the northern and the southern regions. The gas content is considerably less towards the southern region of HD 216658. This may be the reason why relatively less clumping is seen in the southern region of the cavity.   
 \begin{table}
\centering
\caption{The table provides the coordinate offsets from HD 216658, radius and mass of 11 molecular clump-like structures identified in the dendogram analysis.}
\label{tab:2}
\begin{tabular}{ccccc}
\hline
\begin{tabular}[c]{@{}c@{}}Clump\\ Number\end{tabular} & \begin{tabular}[c]{@{}c@{}}R. A offset\\ (arcsec)\end{tabular} & \begin{tabular}[c]{@{}c@{}}Dec offset\\ (arcsec)\end{tabular} & \begin{tabular}[c]{@{}c@{}}Radius\\ (pc)\end{tabular} & \begin{tabular}[c]{@{}c@{}}Mass\\ M\textsubscript{\(\odot\)}\end{tabular} \\ \hline
1 & 476.5 & -164.8 & 0.04 & 19 \\
2 & 564.9 & -105.3 & 0.06 & 85 \\
3 & -309.0 & -11.2 & 0.14 & 3016 \\
4 & -457.6 & -80.7 & 0.05 & 46 \\
5 & 151.6 & 278.9 & 0.05 & 94 \\
6 & 87.9 & 263.6 & 0.03 & 6 \\
7 & -196.2 & 179.0 & 0.13 & 2551 \\
8 & -62.3 & 216.9 & 0.06 & 70 \\
9 & 112.8 & 364.1 & 0.05 & 64 \\
10 & 46.1 & 326.7 & 0.06 & 97 \\
11 & -88.5 & 534.1 & 0.05 & 44 \\ \hline
\end{tabular}
\end{table}

\begin{figure}
   \centering
   \includegraphics[width=1\columnwidth]{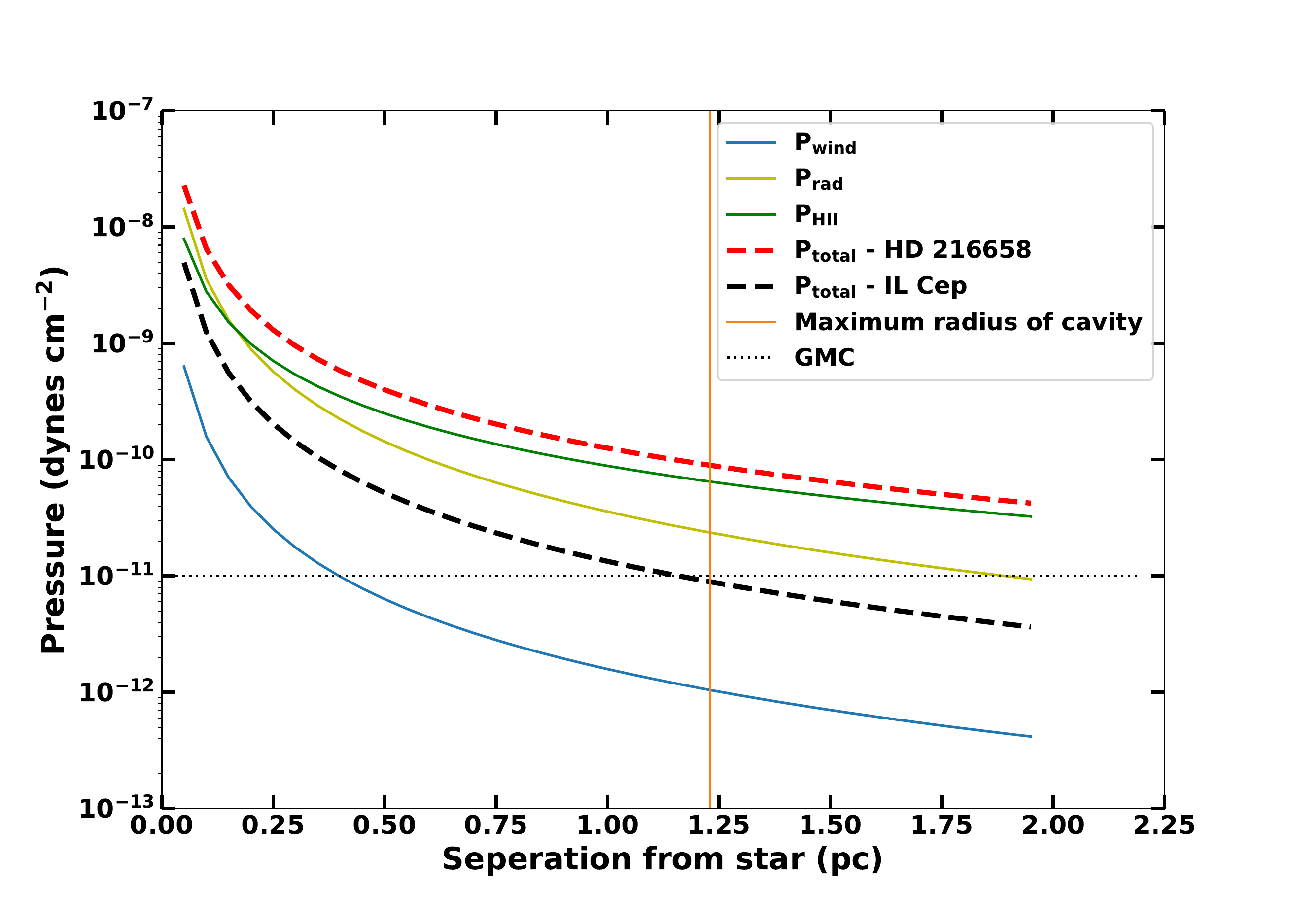}
   \caption{ The figure shows various pressure components produced by HD 216658. The total pressure is shown using the red dashed line. The orange line is the maximum radius of the cavity. The pressure experienced by a GMC is shown by the black dotted line. The total pressure from the system is higher than normal GMC pressure at the maximum radius. The black dashed line shows the total pressure by IL Cep which is an order of magnitude lower than HD 216658. This shows that the HD 216658 itself is capable of producing the "cavity". }
   \label{Fig12}
   \end{figure}

\subsection{Pressure effects due to HD 216658}\label{sec:pressure}

  HD 216658 is the primary source of UV photons in the region that carved out the "cavity". The high mass stars can exert pressure effects due to radiation \citep{Simon2008MNRAS.389.1009S}, wind \citep{Dale2015MNRAS.450.1199D}, and ionized gas \citep{Sternberg2003ApJ...599.1333S} on the neighboring molecular clouds. In this section, we evaluate different pressure components by HD 216658 responsible for creating the cavity. The maximum radius of the cavity around the IL Cep region is 1.2 pc \citep{Zhang2016MNRAS.458.4222Z}. The variation of each of the pressure components from HD 216658 is evaluated up to 2 pc radius and is shown in \autoref{Fig12}. We compute the pressure due to stellar winds using the following equation.
\begin{equation}
    P_{wind} = \frac{M_W V_W}{4\pi D_s^2}
    \label{EQ1}
\end{equation}
where M\textsubscript{W} is the mass-loss rate, V\textsubscript{W} is the stellar wind velocity and D\textsubscript{s} is the distance from the star \citep{Baug2019ApJ...885...68B}. We assumed an average mass loss rate of 10\textsuperscript{-7} M\textsubscript{$\odot$}~yr\textsuperscript{-1} and an average wind velocity of 300 km~s\textsuperscript{-1} for HD 216658, assuming it as a HAeBe star in its PMS evolutionary stage \citep{Strafella1998ApJ...505..299S}.  We estimated the P\textsubscript{wind} extended up to 2 pc using \autoref{EQ1} by varying the D\textsubscript{s} value. The pressure owing to the radiation (P\textsubscript{rad}) is computed using the equation.
\begin{equation}
    P_{rad} = \frac{L_{bol}}{4\pi D_s^2}
    \label{EQ2}
\end{equation}
The bolometric luminosity (L\textsubscript{bol}) of HD 216658 is estimated using V magnitude with necessary bolometric corrections. P\textsubscript{rad} is estimated up to a radius of 2 pc and the pressure curve due to radiation is shown in \autoref{Fig12}. 

Another component affecting the region is the pressure due to the ionized gas (P\textsubscript{HII}). The value of P\textsubscript{HII} can be estimated using the expression given below.

\begin{equation}
    P_{HII} =  \mu_{II} m_H c^{2}_{II} (\frac{S_{Lyc}}{4\pi \beta_2 D_s^3})^{(1/2)}
    \label{EQ3}
\end{equation}

where, the mean molecular weight in an HII region, $\mu$\textsubscript{II}=0.678 \citep{Bisbas2009}, the sound speed in an HII region, c\textsubscript{II}= 11 km~s\textsuperscript{-1}, and the recombination coefficient, $\beta$\textsubscript{2} =2.6 $\times$ 10\textsuperscript{-13} cm\textsuperscript{3} s\textsuperscript{-1}. Possible fluxes of Lyman Continuum Photons that is produced by a typical B0V-type progenitor (S\textsubscript{Lyc} $\sim$ 10\textsuperscript{47.3} photon s\textsuperscript{-1}) is taken from \cite{Panagia1973AJ.....78..929P}. Considering all the parameters, we estimated P\textsubscript{HII} value for IL Cep system upto a radius of 2 pc and is illustrated in the \autoref{Fig12}. The figure is marked with the maximum radius of the cavity (1.2 pc).  Also, the pressure experienced by a cool GMC (Giant Molecular Cloud) (10\textsuperscript{-11} dynes cm\textsuperscript{-2}; \citealp{Baug2019ApJ...885...68B}) is shown in the \autoref{Fig12}. A pressure approximately one order of magnitude higher than 10\textsuperscript{-11} dynes cm\textsuperscript{-2} produced by a source can perturb the region and eventually lead to cavity formation. We apply the similar analysis for the pressure components induced by IL Cep. The calculation is made by keeping wind pressure the same, but the L\textsubscript{bol} and S\textsubscript{Lyc}  are changed for IL Cep accordingly. We find that the total exerted pressure is approximately an order of magnitude lower than HD 216658, and thus cannot be attributable for the observed cavity. The total pressure exerted by IL Cep is also shown in the \autoref{Fig12}.

The analysis shows that HD 216658 system is producing enough pressure to create a cavity of radius 1.2 pc. The calculations are made on the assumption of its PMS nature and the present conditions. The protostellar stage of massive stars can be more violent and energetic, which shapes the expanding envelope around the star \citep{Kuiper2015ApJ...800...86K}. Thus the pressure value is probably underestimated. Considering all the factors it is safe to argue that the cavity is indeed made by HD 216658.    
\begin{figure*}
   \centering
   \includegraphics[width=1\columnwidth]{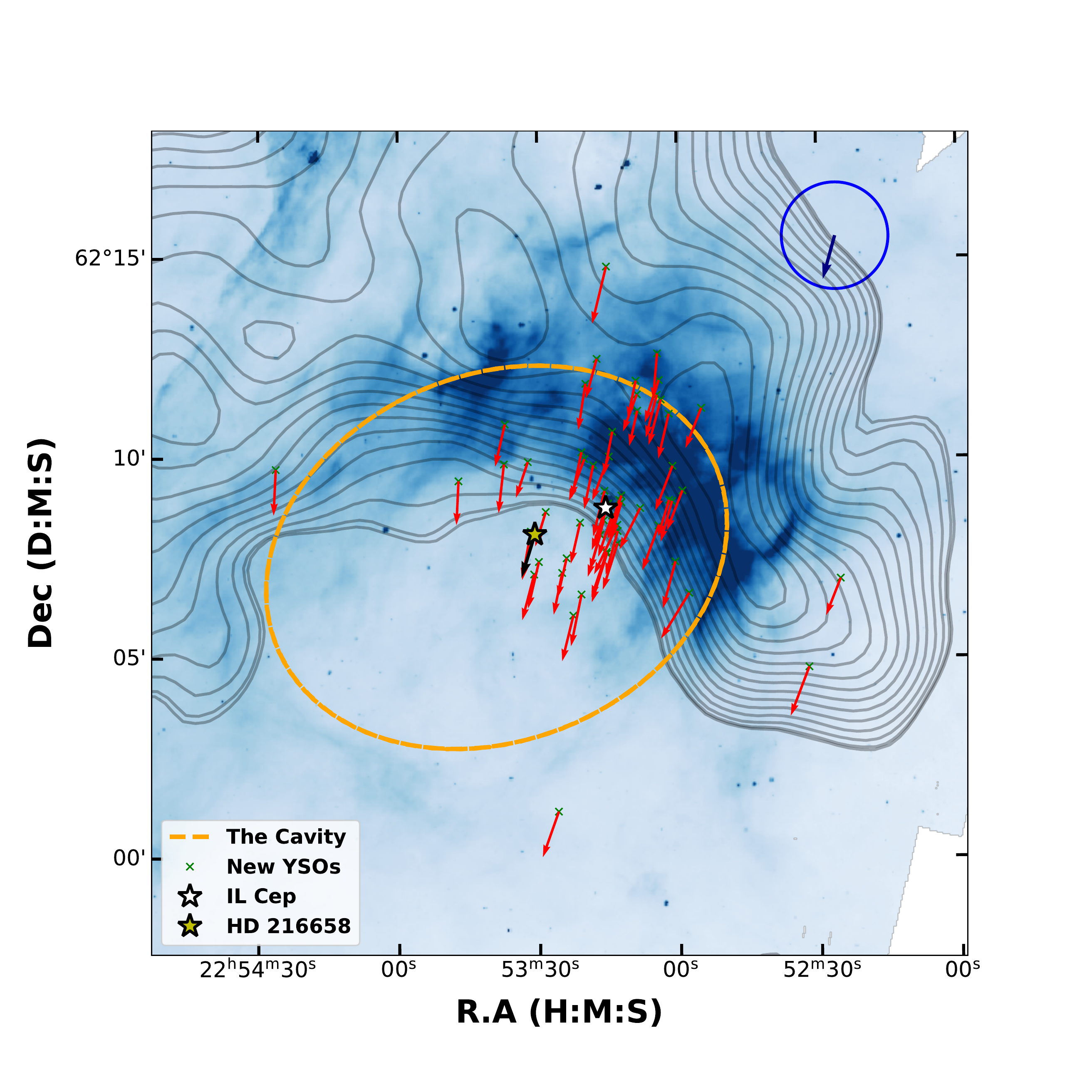}
   \includegraphics[width=1\columnwidth]{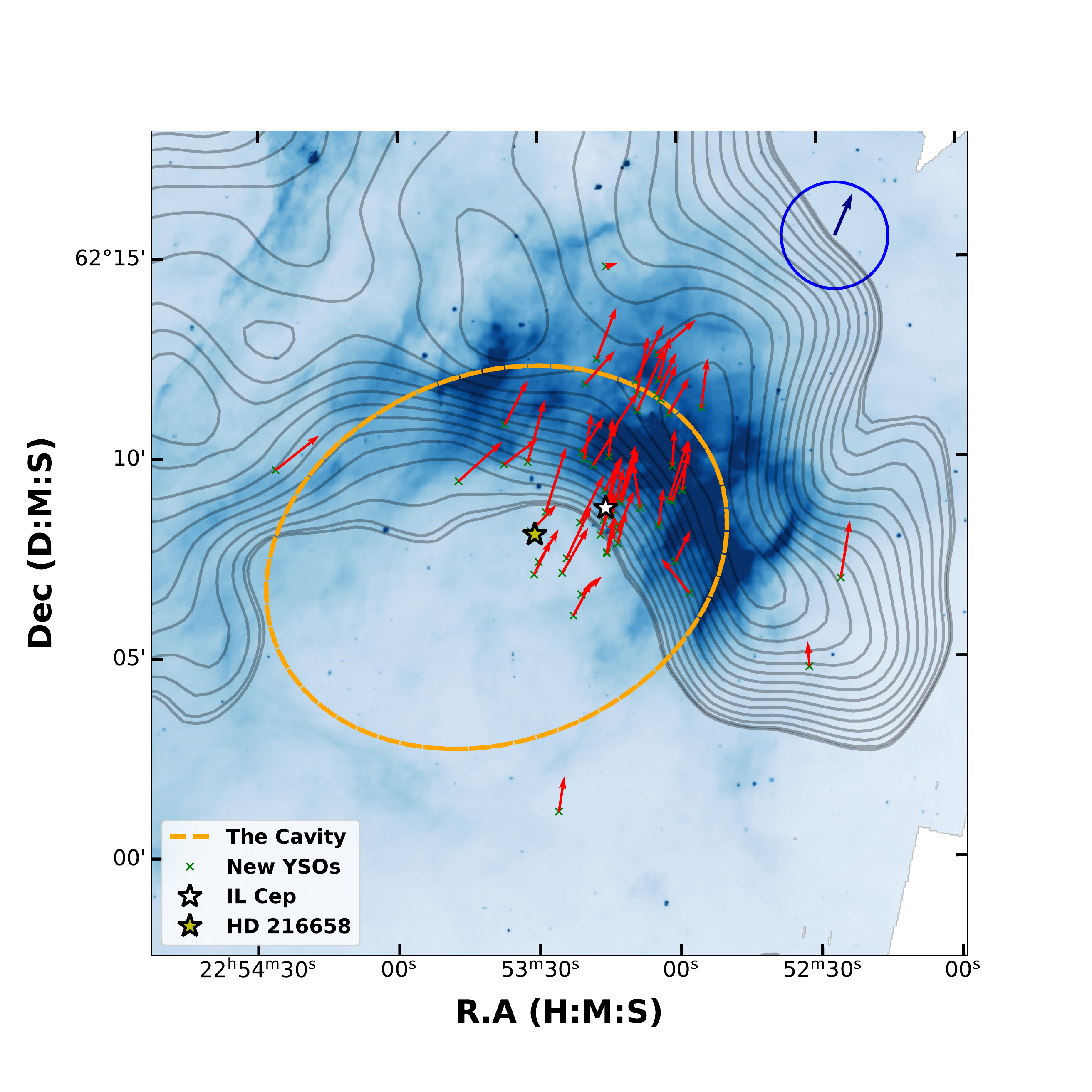}
\caption{Figures show the IRAC 8 $\mu$m image of the IL Cep region with \textsuperscript{12}CO (\textit{J = 1$-$0)} channel map of -6 to -9 km~s\textsuperscript{-1}. The proper motion vectors of the population of stars with a similar transverse velocity of IL Cep are shown in the figures. The left panel figure shows the normal proper motion vectors and the right panel figure has vectors after subtracting the proper motion of HD 216658. The median proper motion of all the stars is shown in the top right of both figures. The reversal of proper motion indicates to rocket effect by HD 216658 on IL Cep and the co-moving stars placed on the expanding cavity. }
   \label{Fig13}
\end{figure*}
   

\subsection{Possible rocket effect by HD 216658 on the IL Cep co-moving association}

The star forming regions are associated with infrared bubbles/cavities or molecular cloud structures \citep{Churchwell2006ApJ...649..759C, Deharveng2010A&A...523A...6D}. These cavity-like structures show an elevated star formation activity than other regions \citep{Zhang2012A&A...544A..11Z, Zhang2013A&A...550A.117Z}. The cavity structures are created due to the pressure (wind \& ionizing radiation) exerted by one or more OB type stars \citep{Bertoldi1989ApJ...346..735B}. The ionization front created by the expansion of the HII regions helps in the formation of bright rims on the boundaries of the cavity \citep{Bedijn1984A&A...135...81B}. These regions get accelerated by a process called "rocket effect" \citep{Oort1955ApJ...121....6O}. The clouds get displaced (cloud shuffling; \citealp{Elmegreen1979ApJ...231..372E}) by this acceleration \citep{McKee2007ARA&A..45..565M}. This displacement due to the rocket effect is prominent on the surface facing the ionizing source \citep{Getman2019MNRAS.487.2977G}. The next generation of stars formed in the accelerating clouds should possess the cloud’s inherent motion, which can be traced by the proper motion of the stars \citep{Dale2015MNRAS.450.1199D}. 

We found that the B0V star HD 216658 is the source of excitation, thereby inducing the formation of a cavity. The Strömgren radius of the star is estimated as 0.5 pc and \autoref{Fig5} shows that the star IL Cep is at the boundary of the Strömgren radius of HD 216658. \cite{Zhang2016MNRAS.458.4222Z} shows that there are bright rims in the vicinity of IL Cep, which indicates a possible action of "rocket effect" on the region where IL Cep is formed. We have analysed the proposition that the star HD 216658 could have triggered a rocket effect on the surrounding region.  IL Cep and the associated co-moving stars may have formed in the expanding cloud inheriting the motion imparted by the rocket effect.


The velocity difference between IL Cep stellar group and HD 216658 population in the sky plane is 3.1 km~s\textsuperscript{-1} with a mean uncertainty of 1.2 km~s\textsuperscript{-1} (see Sect. \ref{sec:two_pop}). The HD 216658 population has a transverse velocity of 11.4 km~s\textsuperscript{-1} and the population is loosely distributed all over the region of study. The HD 216658 population could be part of the bigger parent cloud Cep OB3. This indicates that the IL Cep stellar group with a transverse velocity of 8.3 km~s\textsuperscript{-1} has been affected by a recoil force from the rocket effect, which in turn, reduced its transverse velocity. In other words, the PAH illuminated region (as traced by 8 $\mu m$ IRAC map), where IL Cep is located expands with the plane-of-the-sky velocity of 3 km~s\textsuperscript{-1}.

The argument that the IL Cep stellar group is formed due to the rocket effect caused by HD 216658 is illustrated using the proper motion vectors as well. The star HD 216658 can be considered as belonging to the first generation of stars formed in the region, having the original motion of the parent cloud Cep OB3. We subtracted the proper motion value of HD 216658 from all the stars in the IL Cep stellar group (> 10.3 km~s\textsuperscript{-1}). \autoref{Fig13} shows the 8 $\mu m$ (IRAC 4) image of the region with \textsuperscript{12}CO (\textit{J = 1$-$0)} channel map contours (-6 to -9 km~s\textsuperscript{-1}) from CGPS survey, observed using the FCRAO 14 m telescope. The left panel in \autoref{Fig13} shows the {\textit Gaia} EDR3 proper motion before the subtraction of the proper motion of HD216658 and the right panel shows the proper motion subtracted vectors. The median proper motion vector of the stars is also shown in both figures. We see a reversal of proper motion for most of the co-moving candidates, thus mimicking an expansion. The reversal is more prominent for the stars in the PAH illuminated region, which is reported as the expanding structure. The \textsuperscript{13}CO channel map contours also indicate that the PAH illuminated region is having a velocity of -9 km~s\textsuperscript{-1}.  

It is also worthwhile to note that the radial velocities of IL Cep and HD 216658 are reported as -39.40 $\pm$ 2.00 km~s\textsuperscript{-1} and -27.50 $\pm$ 3.40 km~s\textsuperscript{-1} respectively \citep{Kharchenko2007AN....328..889K}. The minimum difference in the radial velocity of both stars is 6.5 km~s\textsuperscript{-1}. The fact that IL Cep is having a faster radial velocity component than HD 216658 is a clear demonstration of the rocket effect.

We demonstrate using {\textit Gaia} EDR3 data analysis that the Herbig Be star IL Cep and its co-moving companions (IL Cep stellar group) were formed due to the triggered star formation by HD 216658 in the Cep OB3 association. This work may be one of the first observational demonstrations of the phenomenon of the "rocket effect" using astrometric data.

\section{Conclusions}
\label{sect:conclusion}
In this study, we report the clustered star formation inside the "cavity", around the Herbig Be star IL Cep. We used {\textit Gaia} EDR3 astrometric data and other archival photometric surveys to find the accreting PMS stars among the co-moving stars. The major conclusions of the study are given below.

\begin{itemize}
    \item The clustering of stars around IL Cep is identified using the Gaussian fitting of distance and proper motion values. 
    \item The 3 $\times$ WMAD ellipses of d vs $\mu_{\alpha *}$ and d vs $\mu_{\delta}$ is used to find 78 stars that are co-moving with IL Cep.
    
    \item HD 216658, the brightest star that occupies the center of the "cavity" is identified to be the trigger for the formation of "cavity" and the co-moving sources using {\textit Gaia} EDR3 astrometry.
    \item The histogram distribution of the transverse velocities of co-moving stars reveals two populations, each associated with IL Cep and HD 216658. 
    \item The {\textit Gaia} CMD indicates that most of the co-moving stars are coeval with IL Cep. The stars are distributed around 0.1 Myr isochrones. 
    \item The 2MASS and mid-IR CCDm are used to find IR excess candidates among the co-moving stars. We identified 25 stars as Class II sources.
    \item We found that 65 \% of the co-moving stars in the IL Cep stellar group belong to Class III.
    \item From the spectral and SED analysis we found that IL Cep B is a Herbig Ae star of A6 spectral type. This makes IL Cep a visual binary system where both components are HAeBe stars. 
    \item Optical spectrum of HD 216658 shows no emission lines and has a spectral type of B0V. The star appears to be much more evolved than IL Cep and can be perceived as the initial trigger for "cavity" formation.
    \item We found 11 molecular clumps around IL Cep using dendrogram analysis.  
    
    \item The total pressure exerted by HD 216658 on the surrounding region is calculated. The estimated pressure by the star is capable of creating the observed "cavity". 
    
    \item The possibility of formation of IL Cep and co-moving stars by HD 216658 through rocket effect is discussed and demonstrated from the analysis of proper motion vectors.
\end{itemize} 

\section*{Acknowledgements}

We would like to thank the referee for providing helpful comments and suggestions that improved the paper. R.A. thanks Ujjwal and Sudheesh for their valuable suggestions throughout the course of the work and also Newman College, Thodupuzha for providing facilities at the time of the pandemic. This work has made use of data from the European Space Agency (ESA) mission {\textit Gaia} (https://www.cosmos.esa.int/{\textit Gaia}), processed by the {\textit Gaia} Data Processing and Analysis Consortium
(DPAC; https://www.cosmos.esa.int/web/{\textit Gaia}/dpac/
consortium). Funding for the DPAC has been provided by
national institutions, in particular, the institutions participating in
the {\textit Gaia} Multilateral Agreement. Also, we made use of the VizieR catalog
access tool, Simbad and Aladdin, CDS, Strasbourg, France. The research has made use of the NASA/IPAC Infrared Science Archive, which is funded by the National Aeronautics and Space Administration and operated by the California Institute of Technology. 

\section*{Data Availability}

The photometric and astrometric data are publicly available from the VizieR catalog\footnote{https://vizier.u-strasbg.fr/}. The derived data generated in this research and the optical spectra will be shared on a reasonable request to the corresponding author.



\bibliographystyle{mnras}
\bibliography{example} 



\onecolumn

\begin{longtable}[c]{ccccccccc}
\caption{The table provides the relevant data of 78 co-moving stars of IL Cep.}
\label{tab:appendix_1}\\
\hline
\begin{tabular}[c]{@{}c@{}}Gaia EDR3 \\ identifier\end{tabular} & \begin{tabular}[c]{@{}c@{}}Offset from \\ IL Cep \\ (arcsec)\end{tabular} & \begin{tabular}[c]{@{}c@{}}Distance \\ (pc)\end{tabular} & \begin{tabular}[c]{@{}c@{}}$\mu_{\alpha *}$ \\ (mas yr\textsuperscript{-1})\end{tabular} & \begin{tabular}[c]{@{}c@{}}$\mu_{\delta}$ \\ (mas yr\textsuperscript{-1})\end{tabular} & \begin{tabular}[c]{@{}c@{}}G \\ (mag)\end{tabular} & \begin{tabular}[c]{@{}c@{}}G\textsubscript{BP} \\ (mag)\end{tabular} & \begin{tabular}[c]{@{}c@{}}G\textsubscript{RP} \\ (mag)\end{tabular} & Class \\ \hline
\endfirsthead
\endhead
\hline
\endfoot
\endlastfoot
Gaia EDR3 2207203690787004800 & 7.7 & $780^{+7}_{-10}$ & -0.68$\pm$0.02 & -2.13$\pm$0.02 & 11.26 & 12.14 & 10.32 & - \\
Gaia EDR3 2207203626363177344 & 15.1 & $715^{+37}_{-27}$ & -0.68$\pm$0.07 & -2.01$\pm$0.08 & 17.36 & 19.07 & 16.02 & II \\
Gaia EDR3 2207203626363177984 & 18.9 & $788^{+37}_{-33}$ & -0.68$\pm$0.06 & -2.45$\pm$0.05 & 16.46 & 17.72 & 15.36 & III \\
Gaia EDR3 2207203695082652928 & 23.8 & $811^{+33}_{-37}$ & -0.58$\pm$0.06 & -1.86$\pm$0.06 & 17.03 & 19.28 & 15.64 & III \\
Gaia EDR3 2207109854345628544 & 25.2 & $892^{+116}_{-78}$ & -0.58$\pm$0.11 & -1.93$\pm$0.12 & 17.98 & 19.81 & 16.61 & III \\
Gaia EDR3 2207203695077255808 & 25.6 & $822^{+244}_{-147}$ & -0.56$\pm$0.24 & -2.24$\pm$0.24 & 19.19 & 21.16 & 17.69 & - \\
Gaia EDR3 2207203695082652544 & 30 & $783^{+29}_{-22}$ & -0.64$\pm$0.04 & -2.1$\pm$0.05 & 16.44 & 17.8 & 15.18 & II \\
Gaia EDR3 2207109785619441792 & 31 & $718^{+114}_{-82}$ & -1.14$\pm$0.18 & -2.43$\pm$0.18 & 18.68 & 20.58 & 17.24 & II \\
Gaia EDR3 2207109789917705344 & 39.3 & $763^{+40}_{-37}$ & -0.59$\pm$0.07 & -2.14$\pm$0.07 & 17.42 & 18.95 & 16.12 & II \\
Gaia EDR3 2207191875332657664 & 41.2 & $750^{+43}_{-35}$ & -0.62$\pm$0.08 & -2.06$\pm$0.08 & 17.33 & 19.23 & 16.04 & II \\
Gaia EDR3 2207191871033079680 & 44.3 & $792^{+92}_{-67}$ & -0.46$\pm$0.16 & -2.01$\pm$0.16 & 18.65 & 21.06 & 17.27 & II \\
Gaia EDR3 2207109858637178624 & 51.9 & $785^{+95}_{-90}$ & -1.02$\pm$0.16 & -2.03$\pm$0.16 & 18.67 & 20.81 & 17.13 & III \\
Gaia EDR3 2207109785619440000 & 56.2 & $778^{+182}_{-116}$ & -0.72$\pm$0.26 & -2.27$\pm$0.23 & 19.23 & 21.29 & 17.72 & II \\
Gaia EDR3 2207191871033078272 & 66 & $786^{+70}_{-55}$ & -0.76$\pm$0.09 & -2.24$\pm$0.09 & 17.74 & 19.52 & 16.48 & III \\
Gaia EDR3 2207203695082652032 & 66.7 & $761^{+33}_{-32}$ & -0.44$\pm$0.06 & -2.16$\pm$0.06 & 17.21 & 19.03 & 15.89 & - \\
Gaia EDR3 2207098038883614080 & 68.6 & $702^{+130}_{-97}$ & -0.75$\pm$0.25 & -2.38$\pm$0.25 & 19.16 & 21.37 & 17.73 & - \\
Gaia EDR3 2207203896941512960 & 77.1 & $809^{+79}_{-70}$ & -0.82$\pm$0.12 & -2.17$\pm$0.12 & 18.22 & 20.09 & 16.89 & III \\
Gaia EDR3 2207203729442390528 & 79.9 & $774^{+51}_{-34}$ & -0.75$\pm$0.07 & -2.03$\pm$0.07 & 17.36 & 19.12 & 16.06 & II \\
Gaia EDR3 2207109785619439744 & 81.4 & $780^{+49}_{-42}$ & -1.11$\pm$0.08 & -2.93$\pm$0.08 & 17.53 & 19.28 & 16.26 & III \\
Gaia EDR3 2207203656423341440 & 82.8 & $910^{+164}_{-118}$ & -0.57$\pm$0.16 & -2.69$\pm$0.15 & 18.54 & 20.74 & 17.12 & II \\
Gaia EDR3 2207109819979179904 & 83.8 & $707^{+183}_{-93}$ & -0.8$\pm$0.21 & -2.2$\pm$0.21 & 19.07 & 21.19 & 17.62 & II \\
Gaia EDR3 2207203656423343744 & 84.8 & $815^{+51}_{-51}$ & -0.25$\pm$0.1 & -2.8$\pm$0.1 & 17.93 & 19.62 & 16.56 & II \\
Gaia EDR3 2207191905392818816 & 90.1 & $875^{+283}_{-157}$ & -0.52$\pm$0.23 & -1.69$\pm$0.22 & 19.1 & 20.86 & 17.72 & III \\
Gaia EDR3 2207203729442390272 & 91 & $819^{+32}_{-35}$ & -0.48$\pm$0.05 & -2.24$\pm$0.05 & 16.88 & 18.4 & 15.67 & III \\
Gaia EDR3 2207191806607793792 & 95.6 & $864^{+127}_{-100}$ & -0.45$\pm$0.2 & -1.93$\pm$0.18 & 18.55 & - & - & III \\
Gaia EDR3 2207110026137611520 & 96 & $826^{+148}_{-97}$ & -0.54$\pm$0.16 & -1.79$\pm$0.15 & 18.58 & 20.8 & 17.16 & III \\
Gaia EDR3 2207098034588915840 & 99.2 & $766^{+50}_{-39}$ & -1.03$\pm$0.08 & -2.79$\pm$0.08 & 17.53 & 19.07 & 16.33 & III \\
Gaia EDR3 2207191840972921344 & 100 & $829^{+13}_{-13}$ & -1.04$\pm$0.02 & -2.47$\pm$0.02 & 14.72 & 15.72 & 13.73 & III \\
Gaia EDR3 2207110026144274304 & 100.2 & $803^{+672}_{-201}$ & -0.56$\pm$0.33 & -2$\pm$0.32 & 19.72 & 21.3 & 18.32 & III \\
Gaia EDR3 2207203965660990592 & 114.7 & $858^{+251}_{-117}$ & -0.42$\pm$0.2 & -2.14$\pm$0.2 & 18.94 & 20.78 & 17.53 & II \\
Gaia EDR3 2207191840967531904 & 115.7 & $836^{+9}_{-10}$ & -1.55$\pm$0.01 & -2.53$\pm$0.01 & 12.8 & 13.4 & 12 & III \\
Gaia EDR3 2207191600449366784 & 117.4 & $704^{+51}_{-55}$ & -0.42$\pm$0.15 & -2.05$\pm$0.13 & 18.35 & 20.32 & 16.99 & II \\
Gaia EDR3 2207191840974688768 & 117.8 & $732^{+137}_{-102}$ & -0.42$\pm$0.34 & -2.38$\pm$0.32 & 18.39 & - & - & - \\
Gaia EDR3 2207110094866673536 & 118.2 & $808^{+231}_{-128}$ & -0.77$\pm$0.24 & -1.97$\pm$0.24 & 19.22 & 21.23 & 17.71 & III \\
Gaia EDR3 2207110094856981632 & 118.5 & $746^{+52}_{-53}$ & -0.85$\pm$0.11 & -2.21$\pm$0.11 & 17.9 & 20.09 & 16.51 & II \\
Gaia EDR3 2207191634809106816 & 128.3 & $843^{+173}_{-115}$ & -1.09$\pm$0.19 & -2.35$\pm$0.22 & 18.78 & - & - & II \\
Gaia EDR3 2207191630514909184 & 129 & $821^{+71}_{-65}$ & -0.53$\pm$0.13 & -2.28$\pm$0.11 & 18.01 & 20.29 & 16.52 & - \\
Gaia EDR3 2207109755557965952 & 132 & $732^{+31}_{-27}$ & -0.61$\pm$0.06 & -2.31$\pm$0.06 & 17.03 & 18.65 & 15.79 & II \\
Gaia EDR3 2207191596155045504 & 135 & $723^{+70}_{-53}$ & -0.52$\pm$0.11 & -2.54$\pm$0.1 & 17.93 & 19.67 & 16.65 & II \\
Gaia EDR3 2207203759502525056 & 135.5 & $737^{+56}_{-64}$ & -0.58$\pm$0.12 & -1.75$\pm$0.11 & 18.15 & 20 & 16.85 & II \\
Gaia EDR3 2207191634814494336 & 146.7 & $768^{+16}_{-14}$ & -0.58$\pm$0.03 & -2.24$\pm$0.02 & 15.4 & 16.59 & 14.32 & III \\
Gaia EDR3 2207204721575215360 & 153.4 & $889^{+354}_{-203}$ & -0.39$\pm$0.33 & -1.74$\pm$0.34 & 19.47 & 21.29 & 17.94 & II \\
Gaia EDR3 2207192042831672448 & 165.9 & $767^{+65}_{-59}$ & -0.26$\pm$0.11 & -2.39$\pm$0.11 & 18.11 & 20.49 & 16.71 & III \\
Gaia EDR3 2207191531735280128 & 168.7 & $824^{+43}_{-33}$ & -0.55$\pm$0.06 & -2.24$\pm$0.06 & 16.81 & 18.52 & 15.56 & III \\
Gaia EDR3 2207110958148796416 & 170.2 & $740^{+26}_{-24}$ & -0.52$\pm$0.05 & -2.21$\pm$0.05 & 16.6 & 18.14 & 15.4 & II \\
Gaia EDR3 2207110953850443136 & 172.4 & $858^{+242}_{-205}$ & -1.2$\pm$0.33 & -2.34$\pm$0.37 & 19.7 & 21.14 & 18.26 & II \\
Gaia EDR3 2207204721575140864 & 176.7 & $729^{+54}_{-58}$ & -0.68$\pm$0.11 & -1.83$\pm$0.11 & 18.14 & 19.8 & 16.91 & III \\
Gaia EDR3 2207109648180096896 & 178.9 & $724^{+170}_{-141}$ & -1.39$\pm$0.32 & -2.25$\pm$0.42 & 19.58 & 21.21 & 18.01 & II \\
Gaia EDR3 2207110958148795648 & 182.4 & $784^{+15}_{-11}$ & -0.59$\pm$0.02 & -2.23$\pm$0.02 & 14.73 & 15.98 & 13.62 & III \\
Gaia EDR3 2207110953850289920 & 186.3 & $862^{+44}_{-41}$ & -0.54$\pm$0.07 & -2.1$\pm$0.08 & 17.38 & 18.9 & 16.15 & III \\
Gaia EDR3 2207204000020636160 & 188.6 & $709^{+57}_{-54}$ & -0.35$\pm$0.12 & -2.27$\pm$0.12 & 18.01 & 20.28 & 16.55 & II \\
Gaia EDR3 2207204794594275712 & 196.2 & $722^{+34}_{-33}$ & -0.38$\pm$0.07 & -1.86$\pm$0.07 & 17.13 & 18.84 & 15.85 & III \\
Gaia EDR3 2207203798161868288 & 196.3 & $803^{+44}_{-33}$ & -0.47$\pm$0.08 & -2.07$\pm$0.08 & 17.69 & 19.32 & 16.43 & III \\
Gaia EDR3 2207109961716390144 & 203.5 & $821^{+238}_{-163}$ & -1.71$\pm$0.3 & -2.52$\pm$0.32 & 19.61 & 21 & 18.16 & III \\
Gaia EDR3 2207111125648971776 & 207.6 & $792^{+183}_{-124}$ & -0.77$\pm$0.23 & -1.98$\pm$0.24 & 18.97 & 20.96 & 17.54 & III \\
Gaia EDR3 2207110953850296064 & 207.7 & $702^{+77}_{-65}$ & -0.68$\pm$0.16 & -2.07$\pm$0.17 & 18.72 & 21.06 & 17.27 & - \\
Gaia EDR3 2207203866881344512 & 214.4 & $840^{+64}_{-71}$ & -0.83$\pm$0.12 & -2.92$\pm$0.11 & 18.14 & 20.12 & 16.76 & III \\
Gaia EDR3 2207109686834905856 & 216.7 & $833^{+96}_{-93}$ & -0.33$\pm$0.14 & -2.87$\pm$0.15 & 18.51 & - & - & III \\
Gaia EDR3 2207190771521448960 & 220.7 & $868^{+433}_{-197}$ & -0.84$\pm$0.44 & -2.95$\pm$0.39 & 19.94 & 21.45 & 18.49 & III \\
Gaia EDR3 2207204755934881536 & 224 & $899^{+200}_{-111}$ & -0.53$\pm$0.18 & -1.94$\pm$0.18 & 18.81 & 20.84 & 17.42 & II \\
Gaia EDR3 2207192356368997888 & 224.4 & $823^{+32}_{-32}$ & -0.1$\pm$0.05 & -2.15$\pm$0.05 & 16.59 & 17.93 & 15.45 & III \\
Gaia EDR3 2207205000752704384 & 244.4 & $737^{+19}_{-14}$ & -0.19$\pm$0.04 & -2.26$\pm$0.04 & 15.73 & 17.08 & 14.54 & III \\
Gaia EDR3 2207192081491094144 & 249.4 & $839^{+59}_{-43}$ & -0.99$\pm$0.07 & -2.53$\pm$0.06 & 17.32 & 18.84 & 16.16 & III \\
Gaia EDR3 2207096900724266752 & 266.7 & $822^{+289}_{-303}$ & -0.74$\pm$0.75 & -3.11$\pm$0.75 & 20.41 & 22.12 & 18.96 & III \\
Gaia EDR3 2207192352069321728 & 269.2 & $817^{+11}_{-11}$ & -1.45$\pm$0.02 & -2.51$\pm$0.02 & 14.57 & 15.67 & 13.51 & III \\
Gaia EDR3 2207192425088470272 & 291.5 & $815^{+7}_{-8}$ & -1.08$\pm$0.01 & -2.57$\pm$0.01 & 13.74 & 14.77 & 12.72 & III \\
Gaia EDR3 2207192425083076480 & 292.1 & $874^{+114}_{-82}$ & -0.89$\pm$0.15 & -2.78$\pm$0.14 & 18.19 & 20.02 & 16.81 & III \\
Gaia EDR3 2207109510741529216 & 295.7 & $855^{+267}_{-169}$ & -1.24$\pm$0.4 & -2.51$\pm$0.52 & 19.98 & 21.38 & 18.49 & III \\
Gaia EDR3 2207204893373835904 & 301.1 & $777^{+200}_{-160}$ & -0.29$\pm$0.4 & -2.91$\pm$0.33 & 19.49 & 21.79 & 17.96 & II \\
Gaia EDR3 2207192115850832768 & 340.9 & $804^{+11}_{-14}$ & -1.15$\pm$0.02 & -2.59$\pm$0.02 & 15.16 & 16.19 & 14.15 & - \\
Gaia EDR3 2207205103831917312 & 362.1 & $731^{+145}_{-109}$ & -0.68$\pm$0.22 & -2.8$\pm$0.22 & 18.95 & 21.29 & 17.34 & III \\
Gaia EDR3 2207110301015137536 & 368.1 & $823^{+227}_{-133}$ & -0.72$\pm$0.24 & -1.84$\pm$0.27 & 19.3 & 20.88 & 17.93 & III \\
Gaia EDR3 2207205168251744640 & 383.9 & $803^{+194}_{-100}$ & -1.62$\pm$0.17 & -3.22$\pm$0.17 & 18.37 & 20.51 & 16.88 & III \\
Gaia EDR3 2207109270222912512 & 387.4 & $771^{+53}_{-45}$ & -0.92$\pm$0.09 & -2.42$\pm$0.08 & 17.34 & 19.17 & 16 & III \\
Gaia EDR3 2207187541706040832 & 416.9 & $823^{+322}_{-221}$ & -1.81$\pm$0.38 & -2.46$\pm$0.32 & 19.74 & 21.37 & 18.34 & III \\
Gaia EDR3 2207205550508517760 & 427.7 & $804^{+19}_{-24}$ & -1.19$\pm$0.04 & -2.8$\pm$0.04 & 15 & 17.49 & 13.57 & III \\
Gaia EDR3 2207093572121225728 & 461.2 & $837^{+64}_{-68}$ & -0.79$\pm$0.12 & -2.23$\pm$0.09 & 17.58 & 19.65 & 16.22 & III \\
Gaia EDR3 2207192901825130752 & 498.3 & $882^{+56}_{-58}$ & -0.1$\pm$0.09 & -2.24$\pm$0.08 & 17.61 & 19.28 & 16.35 & III \\ \hline
\end{longtable}

\bsp	
\label{lastpage}
\end{document}